\documentclass{amsart}       
\usepackage{graphicx}
\usepackage{newtxtext,newtxmath}
\usepackage{hyperref}
\usepackage{color}
\usepackage{xcolor}

\definecolor{darkblue}{RGB}{63,101,134}
\hypersetup{
    colorlinks,
    linkcolor={black},
    citecolor={darkblue},
    urlcolor={darkblue}
}
\usepackage{todonotes}

\usepackage{fancyvrb}
\fvset{formatcom=\color{red}}
\fvset{listparameters={\setlength{\topsep}{0pt}\setlength{\partopsep}{0pt}}} 

\newtheorem{theorem}{Theorem}[section]
\newtheorem{proposition}[theorem]{Proposition}

\theoremstyle{definition}
\newtheorem{definition}[theorem]{Definition}
\newtheorem{example}{Example}
\begin{document}

\title{Linear Differential Equations as a Data-Structure
}
\thanks{This
work has been supported in part by FastRelax ANR-14-CE25-0018-01.}
\author{Bruno Salvy}
\address{
              INRIA, Laboratoire LIP, Universit\'e de Lyon, CNRS, ENS
              Lyon, UCBL, France}
              \email{Bruno.Salvy@inria.fr}


\begin{abstract}
A lot of information concerning solutions of linear
differential
equations can be computed directly from the equation. It is therefore
natural to consider these equations as a data-structure, from which
mathematical properties can be computed. A variety of algorithms has
thus been designed in recent years that do not aim at ``solving'', but
at computing with this representation. Many of these results are
surveyed here.
\end{abstract}
\maketitle

\section{Introduction}
\label{intro}
Computer algebra is a subfield of ``foundations of
computational mathematics'' devoted to exact
mathematical objects: their 
effectivity (what can be computed or decided?) and their complexity 
(how fast?). The first conference I am aware of that was devoted
purely to symbolic and algebraic computation was held in Washington
in~1966. Since then, for more than 50 years, numerous algorithms have
been developed, many of which are available in today's popular
computer algebra systems. This article presents a small fraction of
the recent work in this area dedicated to linear differential
equations and biased towards my own interests. It is mostly based on
an invited talk at FoCM'17. The choice
of presentation is to outline
the underlying ideas through simple examples or algorithms and not
put too much stress on proofs or general or formal statements, for
which pointers to references are given.

\bigskip

There are several motivations for exact computations with linear
differential equations, depending on the origin of these equations.
\paragraph{Special functions.}
Many classical elementary or special functions are solutions of linear
differential equations. This includes exponential, logarithm, rational
functions, hypergeometric functions or generalized hypergeometric
functions in their many variants (Bessel functions, Airy functions,
Struve functions,\dots), orthogonal polynomials, etc. In this case,
the differential equations have small order and the questions are to
derive automatically formulas that practitioners currently look up in
dedicated encyclopedias~%
\cite{AbramowitzStegun1992,PrudnikovBrychkovMarichev1986a,OlverLozierBoisvertClark2010}. 

\paragraph{Generating functions.}
Another
source of linear differential equations is provided by generating
functions in combinatorics. There, the equations annihilate a power
series whose $n$th coefficient counts the number of objects of
interest of size~$n$. The mere knowledge that this power series
satisfies a linear differential equation gives information on the
possible asymptotic behaviour of those coefficients. From the actual
differential equation one can often derive precise asymptotics. 
In this area, the linear differential equations
are often of high order. Their computation itself is
difficult and requires efficient dedicated algorithms. A spectacular
recent example was the study of so-called
\emph{Gessel walks} by Alin Bostan and Manuel 
Kauers~\cite{BostanKauers2010}. These are walks confined to~$
\mathbb{N}^2$,
starting from the origin and with steps restricted to~$\{(-1,0),
(-1,-1),(1,1),(1,0)\}$. The coefficient of~$t^n$ in the generating
function is a polynomial in two extra variables~$x$ and~$y$, where the
coefficient of~$x^iy^j$ is the number of such walks of length~$n$
ending at the point with coordinates~$(i,j)$. In an intermediate step
of their proof that this generating function is algebraic, they
construct a linear differential equation of order~11 with coefficients
that are polynomials of degree up to~96 in~$t$ and 78 in~$x$ and
integer coefficients of up to 61 decimal digits. This is only for the
value at~$y=0$ of the generating function! Such a computation would be
impossible with straightforward algorithms.

\paragraph{Periods.}
Linear differential equations of potentially high order also arise in
more geometric contexts. The
integral of a rational
function in $n+1$ variables over a cycle in~$\mathbb{C}^n$ satisfies a
linear differential equation in the remaining variable called a 
\emph{Picard-Fuchs equation}. Algebraic integrands can also be
allowed without changing the class of integrals, since algebraic
functions can be expressed as residues of rational functions~\cite{DenefLipshitz1987}. An
early
example of a linear differential equation arising in this way is
Euler's
computation of the perimeter of an ellipse as a function of its
eccentricity. More recently, the computation of differential
equations of this type has given rise to efficient algorithms for the
computation of multiple binomial sums (see \S~\ref{sec:CTnew})
and volumes of semi-algebraic sets~\cite{LairezSafey-El-Din2018}.

\paragraph{}
The following two simple definitions make many statements more
compact and set the notation for the sequel. There, as in the rest of
this article, $\mathbb{K}$
denotes an arbitrary field of
characteristic~0, even though some of the statements hold more
generally.
\begin{definition}
	A power series~$S(z)\in\mathbb{K}[[z]]$ is called 
	\emph{differentially finite}, or in short,
	D-finite, when there
	exist polynomials~$p_0(z),\dots,p_m(z)$ in~$\mathbb{K}[z]$ with
	$p_m\neq0$
	such that
\begin{equation}\label{eq:lindeq}
		p_m(z)S^{(r)}(z)+\dots+p_0(z)S(z)=0.
\end{equation}
\end{definition}
\begin{definition}
	A sequence~$(u_n)$ of elements of~$\mathbb{K}$ is called 
	\emph{polynomially recursive}, or in short,
	P-recursive, when
	there exist polynomials~$a_0(n),\dots,a_r(n)$ in~$\mathbb{K}[n]$
	with $a_r\neq0$ such that
\begin{equation}\label{eq:linrec}
a_r(n)u_{n+r}+\dots+a_0(n)u_n=0,\qquad\text{for all
	$n\in\mathbb{N}$}.
\end{equation}
\end{definition}
A classical important observation relates these two
families.
\begin{proposition}The power series $S(z)=\sum_{n\ge0}{u_nz^n}\in
\mathbb{K}[
[z]]$ is
differentially finite if and only if the sequence $(u_n)$ is
polynomially recursive.
\end{proposition}
The computation of the recurrence from the differential equation or
conversely are straightforward. (An efficient algorithm is known for
 large orders and degrees~%
\cite{Bostan2003,BostanChyzakGiustiLebretonLecerfSalvySchost2017}.)
Even such a simple proposition has nontrivial computational
consequences.
\begin{example}
In order to compute the coefficient of $X^N$ in a high power like $P=
(1+X)^N(1+X+X^2)^N$, an efficient method starts from the first-order
linear differential equation satisfied by this polynomial:
\[\frac{P'}{P}=\frac{N}{1+X}+\frac{N(2X+1)}{1+X+X^2}.\]
From there, the Proposition asserts the existence of a linear
recurrence (of order~3
with coefficients of degree~1) for the coefficients of~$P$. Using this
recurrence makes it possible to obtain the $N$th coefficient
efficiently, without computing the previous ones, by the methods of
Section~\ref{sec:binsplit}.

The same reasoning extends to high-order coefficients of high-order
powers of arbitrary polynomials, since the
polynomial~$P^k$ satisfies the linear
differential equation $Py'-kP'y=0$, which is of order~1 with
coefficients of degree at most~$\deg P$,
leading to a linear recurrence of order~$\deg P$ with
coefficients of
degree~1.
\end{example}

\paragraph{}
\begin{figure}
\centerline{\includegraphics[width=\textwidth]{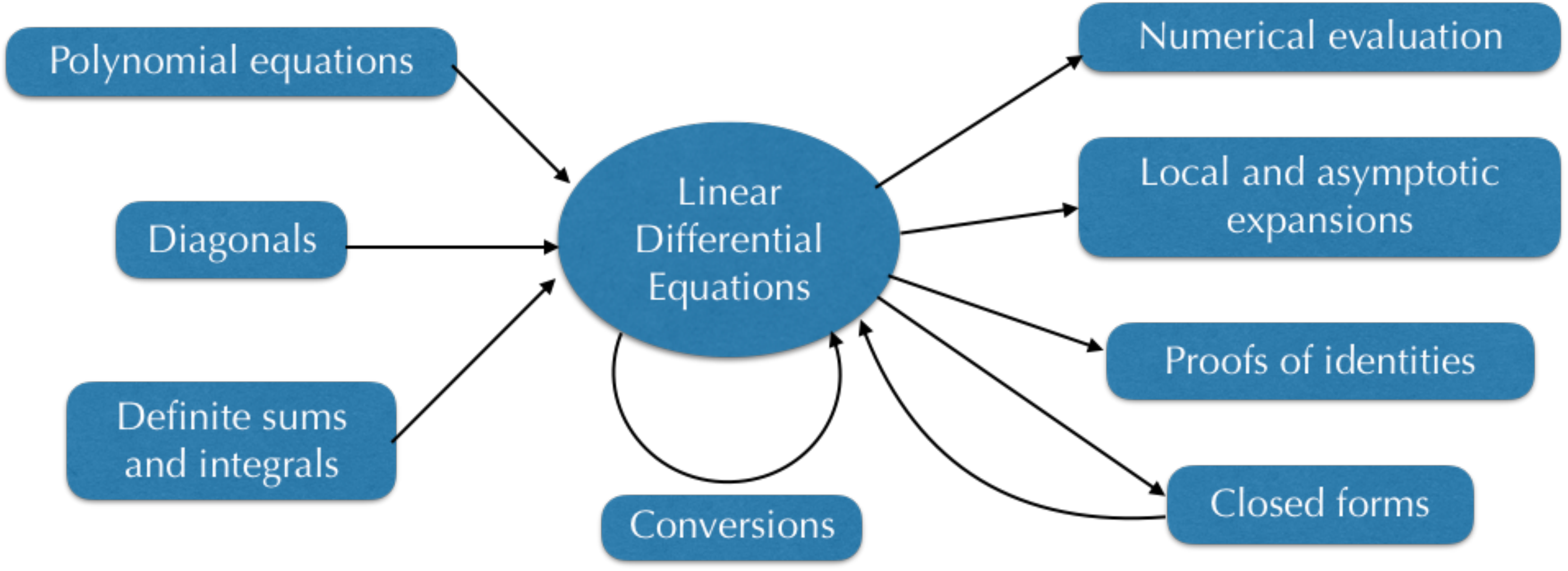}}
\caption{Plan of the article\label{fig:plan}}
\end{figure}
The plan of this article consists in visiting Figure~\ref{fig:plan}
from
right to left.
The central point is that linear
differential equations with polynomial coefficients provide a useful
representation for their solutions,
even when the order or the degree of the
coefficients of the equation are large. From the equation and
its initial conditions a lot of information concerning the solution
can be computed exactly and often efficiently as well. This is 
covered in Part~I. An important part of computer algebra
that is not discussed here is the computation of closed-form
solutions of these equations using differential Galois theory~%
\cite{PutSinger2003}. An advantage of having solutions in closed form
is that these formula provide analytic continuation ``for free''.
However, even when closed-forms are available,
which is rare, they are often not so appropriate for computations.
Our approach will be to convert them into a linear differential
equation, to which the algorithms described here apply.
Once it is clear that many operations can be performed efficiently on
linear differential equations, a natural objective is to design
algorithms that compute such equations to solve other problems. This
is the topic of Part~III where differential equations are
computed for algebraic functions, for multiple integrals and for
generating functions of sums. 

\part*{I. Using Linear Differential Equations Exactly}

\section{Numerical values from linear recurrences}\label{sec:binsplit}
Numerical values can be considered as exact mathematical objects
when a bound on the
approximation
error is known and can be made arbitrarily small. It turns out that
this can be achieved for all
solutions of linear differential equations, with a very good
complexity with respect to the desired precision, by exploiting linear
recurrences and using only elementary ideas.

\subsection{Fast multiplication}
In terms of complexity, the starting point is the Fast Fourier
Transform (FFT). The
theoretical complexity for multiplying two $n$-digit integers is~$O
(n\log
n\log\log n)$ bit operations, with recent improvements~\cite{Furer2009,HarveyHoevenLecerf2016}
decreasing this
bound slightly. We use the notation~$\tilde{O}(n)$ for such
complexities, meaning that
they are in~$O(n\log^k n)$ for some~$k$. More generally, $\tilde{O}
(f(n))$ for a function~$f$ tending to~$+\infty$ means~$O(f(n)\log^k
f(n))$ for some~$k>0$. We say that an algorithm is 
\emph{quasi-optimal} when its complexity is $\tilde{O}(n)$, for $n$
the sum of the sizes of its input and output.

In practice, two integers of a million decimal digits
can be multiplied in much less than one second on current laptops.
Using Newton iteration, that same complexity of~$\tilde{O}(n)$ and
similar timings are
reached for the computation of $n$ digits of reciprocals, square-roots
and many
other operations~\cite{Brent1976a}.

\subsection{Efficient computation of \texorpdfstring{$n!$}{n!}} Fast
multiplication alone
is
not sufficient to compute $n!$ fast if one uses it naively. By
Stirling's formula, the bit size of $k!$ grows roughly like $k\log k$,
so that computing $n!$ as $((1\times 2)\times 3)\dotsm$ would
lead to a complexity in~$\tilde{O}(n^2)$, even if FFT is used. What
happens is that all $k!$ for $k=1,\dots,n$ are obtained during
intermediate computations and since the total bit size of those
is~$\tilde{O}(n^2)$, a lower bound in~$n^2$ is unavoidable. 

However, $n!$ can be computed more efficiently by a divide-and-conquer
approach, using the equation
\[n!=\underbrace{n\times\dots\times\lceil n/2\rceil}_{\text{size
$O(n\log n)$}}\times\underbrace{(\lceil n/2\rceil-1)\times\dots\times
1}_{\text{size
$O(n\log n)$}}
.\]
By Stirling's formula, each half product has size growing
asymptotically like~$\frac12n\log n$, so that their product can be
computed in~$\tilde{O}(n)$ bit operations. Applying the same
divide-and-conquer approach recursively leads to a so-called `product
tree', whose complete computation is
performed in~$\tilde{O}(n)$ bit operations~\cite{Bernstein2008a}. For the special case
of~$n!$, it is even possible to save some of the logarithms hidden in
the~$\tilde{O}$ notation by looking at prime factors of~$n$~
\cite{Borwein1985},
but this idea does not generalize as much as the product-tree technique.

\subsection{Binary splitting} The computation of $n!$ above does not
make use of commutativity and thus extends to the efficient
computation of products of  matrices of integers. Rewriting a linear
recurrence of order $k$ over scalars into a first-order linear
recurrence over vectors of dimension~$k$ therefore extends this method
to arbitrary linear recurrences.
\begin{example}\label{ex:sum-exp}
The sequence
\begin{equation}\label{eq:sum-exp}
e_n=\sum_{k=0}^n\frac1{k!},
\end{equation}
is easily seen to satisfy the second-order linear recurrence~$e_n=
\frac{1}{n}((n+1)e_
{n-1}-e_{n-2})$, $n\ge2$, or equivalently
\[\begin{pmatrix}e_n\\ e_{n-1}\end{pmatrix}=\frac1n
\underbrace{\begin{pmatrix}n+1&-1\\n&0\end{pmatrix}}_{A(n)}
\begin{pmatrix}e_{n-1}\\ e_{n-2}\end{pmatrix}.\]
Using the initial conditions leads to
\[\begin{pmatrix}e_n\\ e_{n-1}\end{pmatrix}=
\frac1{n!}{A!}(n)\begin{pmatrix}1\\0\end{pmatrix},\]
where $A!(n)$ denotes the \emph{matrix factorial}  $A
(n)A(n-1)\dotsm A(1)$. This product is computed as above by a
divide-and-conquer method, which gives the $n$th element $e_n$ in~$
\tilde{O}(n)$ bit operations, i.e., in a quasi-optimal way~%
\cite{Brent1976}.
\end{example}
This reasoning leads to the following useful result.
\begin{theorem}\cite[Thm.~6.1]{ChudnovskyChudnovsky1988}
If the sequence~$(u_n)$ is given by a linear recurrence with polynomial coefficients in~$\mathbb{Q}
[n]$ and initial conditions in~$\mathbb{Q}$, all numerators and
denominators of the rational numbers occurring in the initial
conditions and in the coefficients of the recurrence being bounded by
a fixed~$K$, then as $N\rightarrow\infty$, the~$N$th element~$u_N$ is
a rational number whose
numerator and denominator have bit size bounded by~$O(N\log N)$ and
can be
computed
in $O(N\log^3N)$ bit operations. The result also holds for initial
conditions
as large as $O(N\log N)$ bits.
\end{theorem}
Note that in the worst case, this computation is much faster than
simply writing down all of $u_0,\dots,u_N$ (not to mention their
computation), which would require a number of bits of order $N^2\log
N$.

This theorem gives the complexity of computing the value~$u_N$ as an
unreduced rational number.
If it is necessary to reduce the result to lowest terms, the final gcd
between numerator and denominator and subsequent divisions also fit
within this complexity bound using a fast algorithm for the gcd.
If what is needed is not a rational number but a numerical estimate,
then by classical techniques based on Newton iteration, one can also
obtain as many as~$O(N\log^2 N/\log\log N)$ digits of the decimal
expansion
within the same
complexity bound.

A more precise estimate of the size and complexity in this theorem,
taking into account the degree of the polynomial coefficients of the
recurrence, the bound~$K$ on the integers and the order of
the recurrence can be obtained without any extra difficulty~\cite[chap.~15]
{BostanChyzakGiustiLebretonLecerfSalvySchost2017}.
This method is very powerful and much more complicated sums than the
truncation~\eqref{eq:sum-exp} of~$\exp(1)$ can be computed efficiently
that way. 
\begin{example}In particular, all recent record computations of $\pi$
use
the following formula discovered in~1989 by the Chudnovsky's~\cite{ChudnovskyChudnovsky1989}:
\[\frac1\pi=\frac{12}{C^{3/2}}\sum_{n=0}^\infty{\frac{(-1)^n(6n)!
(A+nB)}{(3n)!n!^3C^{3n}}},\]
with $A=13591409$, $B=545140134$ and $C=640320$. This series gives
roughly 14~digits per term. That observation alone is not
sufficient to yield a fast algorithm, which is obtained by observing
that the summands satisfy a linear recurrence of order~1 which can be
subjected to binary splitting. (The final division and square-root
are handled by Newton iteration.) In
theory, the techniques based on the arithmetic-geometric mean give an
algorithm that is faster by a factor of~$\log N$ for the computation
of $N$ decimal digits, but that method is more delicate to implement
and
 thus binary splitting is preferred, even for record computations~%
\cite{BorweinBorwein1987,HaiblePapanikolaou1998,Bellard2010}.
\end{example}
\section{Numerical values from linear differential equations}
As the example above suggests, this efficient method for
computing the $N$th element of polynomially recursive sequences
extends to give a fast algorithm
for the numerical evaluation of differentially finite functions. If
$f$ is differentially finite, 
$(f_m)$ are the coefficients of its Taylor expansion at the origin
 and~$x$ is a rational number inside
the disk of convergence of~$f$,
then the value of~$f(x)$ is the limit of the sequence
\[F_n(x)=\sum_{m=0}^n{f_mx^m},\qquad n\rightarrow\infty.\]
From a linear recurrence of order $k$ for~$(f_m)$, one deduces a
linear recurrence of order~$k+1$ for $F_n(x)$, whose $n$th element can
be computed efficiently using a product tree for the the matrix
 factorial as above~\cite{BeelerGosperSchroeppel1972,ChudnovskyChudnovsky1988}. 
Example~\ref{ex:sum-exp} illustrates this idea on the differential
equation~$y'-y=0$ with $y(0)=1$ that, in our context, defines the
exponential.

Rough estimates show that in all cases, the tail of the power series
$\sum_{m>n}
{f_mx^m}$ decreases sufficiently fast for $O(n)$ terms to be sufficient
for the computation of $n$ digits of $f(x)$. In order to deduce from this method an algorithm for numerical
evaluation, it is thus sufficient to provide effective bounds on
that tail. This can be achieved by using the linear recurrence on
the coefficients~$(f_m)$ to produce a majorant series whose speed of
convergence is
under control~\cite{Hoeven1999,MezzarobbaSalvy2010}.

\subsection{Analytic continuation} The same approach that
gives
arbitrarily precise estimates for the value of a differentially
finite power series at a rational point inside its disk
of convergence also applies to the case of a complex point with
rational real and imaginary parts. It also applies to the first
derivatives of the power series at such a point. 
Thus one can compute arbitrarily precise initial conditions for the
same differential equation translated at such a point. From there, applying the same
process again makes it possible to compute numerical approximations
at any point given by a polygonal path starting from
the origin, using only points with (preferably small) rational
coordinates as vertices and
avoiding the (finitely many) singularities of the equation. This
method produces numerical evaluation at precision~$N$ in quasi-optimal
complexity~$\tilde{O}(N)$. Again, the whole computation only involves
rational numbers and no round-off errors occur.

\begin{figure}
\centerline{\includegraphics[height=4cm]{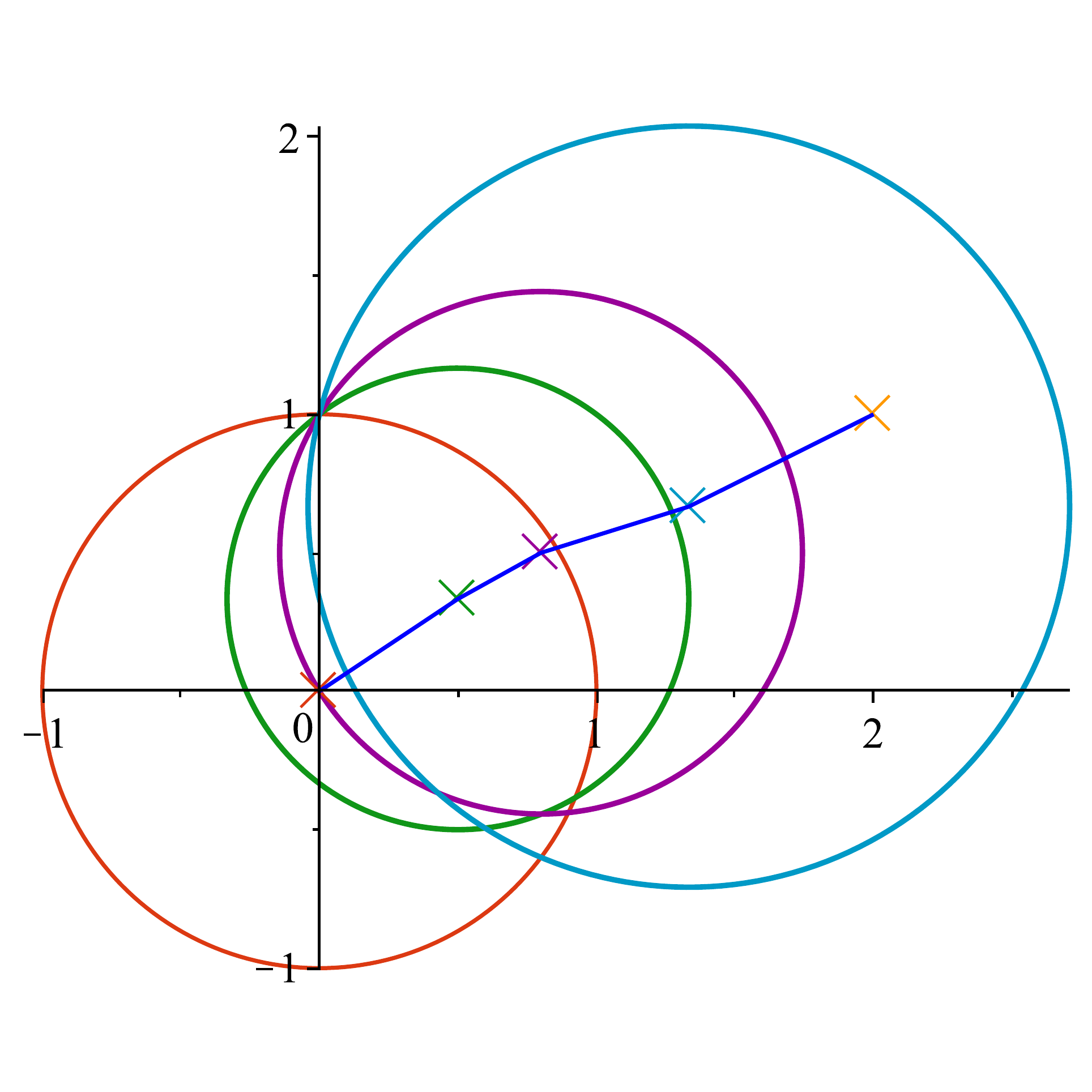}}
\caption{Analytic continuation of arctan from 0 to $2+i$, using
automatically selected intermediate
points with small bit size\label{fig:plot_arctan} inside the disk of
convergence centered at the previous point. The circles of
convergence of the successive power series are given, with the same
color as their center.}
\end{figure}

Low complexity relies on a precise control over the integers
occurring in intermediate computations. When the differential equation
is translated at a point with large rational real or imaginary part,
then the
linear recurrence that results inherits large rational coefficients
that weigh on its evaluation. If the point where the evaluation is
required itself has small rational real and imaginary parts, then it
is always possible to find intermediate points of the same kind in the
analytic continuation path and the complexity remains moderate.
\begin{example}Figure~%
\ref{fig:plot_arctan} displays the domains of convergence of the
series obtained
at the intermediate points taken by M.~Mezzarobba's
\texttt{ore\_algebra\_analytic} package~\cite{Mezzarobba2016} to
evaluate $\arctan(2+i)$ starting from the origin, using this strategy
with further refinements regarding the choice of intermediate
points so that their bit size remains small.
\end{example}

\subsection{Bit burst}
When
the targeted evaluation point is not a rational number but is known
only via an approximation (e.g., $\pi$) then one can use analytic
continuation again. Even if
the point is inside the disk of convergence, this makes it possible to
trade integer size for number of terms in the power series by a
technique called \emph{bit
burst}~\cite{ChudnovskyChudnovsky1988}. For instance, in order
to evaluate  at $\pi$ a function given by its differential
equation and initial
conditions using this
method, one would use as intermediate points the first rational
numbers in the sequence
$\left(\lfloor 2^
{2^i}\pi\rfloor 2^{-2^i}\right)_{i\ge0}$.
While the size of the numerators and denominators of these rational
numbers grows with~$i$, the number of terms of the power series needed
to
obtain the desired accuracy decreases.
These results
are summarized in the following theorem.
\begin{theorem}\cite[Thm.~5.2]{ChudnovskyChudnovsky1988}
If the power series~$y(z)$ is given by a linear differential
equation
with polynomial coefficients in~$\mathbb{Q}[z]$ and initial
conditions, all numerators and denominators of the rational
numbers occurring in the coefficients of
the equation being bounded by $10^{K}$, and all initial conditions
being given at precision~$10^{-K}$, then given a point~$\zeta$
inside
the disk of convergence of~$y(z)$ at precision~$10^{-K}$,  the
value of~$y(\zeta)$ at precision~$10^{-K}$ can be evaluated
in~$\tilde{O}(K)$ bit operations.
\end{theorem}
More precise estimates can be derived in terms of all the parameters,
with refinements for special cases and generalizations to singular
 points%
~\cite{ChudnovskyChudnovsky1988,ChudnovskyChudnovsky1990,Hoeven2001,Hoeven2007b,Mezzarobba2010,BostanChyzakGiustiLebretonLecerfSalvySchost2017}.

\section{Local and asymptotic expansions}

By the Picard-Lindel\"of theorem (that we call Cauchy-Lipschitz in
France), the linear differential equation~\eqref{eq:lindeq} admits a
basis of \emph{analytic} solutions in the neighborhood of any point
that is not a zero of its leading coefficient~$p_m(z)$. For those
solutions, Taylor expansions can be computed to arbitrary order efficiently using the linear recurrence that the coefficients satisfy.

\subsection{Singular behavior}
In a neighborhood of a zero~$a$ of the leading coefficient, the
Picard-Lindel\"of theorem does not hold and the equation
may present singular solutions. 
A classification of the possible
behaviors of solutions is known. An important part is played by the
\emph{indicial polynomial} of the equation at~$a$. This polynomial
in~$\mathbb{K}(a)[s]$ is
obtained as the leading coefficient of the power series obtained by
evaluating the linear differential equation at~$(x-a)^s$ for a
formal~$s$ and multiplying by~$(x-a)^{-s}$. It is equal, up to an
integer shift of~$s$, to the leading coefficient of the recurrence
satisfied by power series solutions of the differential equation
at~$a$. In the case of an \emph{ordinary point}, i.e., when $p_m
(a)\neq0$, the indicial polynomial is simply~$s(s-1)\dotsm(s-m+1)$.
More generally, when the degree of the indicial polynomial at~$a$ is
equal to the order of the differential equation, the point $a$ is
called a \emph{regular singular point} or a \emph{Fuchsian}
singularity. It is called an
\emph{irregular singular point} otherwise.
\begin{theorem}\cite{Fabry1885} 
If~$a$ is a regular singular point, then Eq.~\eqref{eq:lindeq} admits
a
basis of formal solutions of the form
\begin{equation}\label{eq:fuchs}
(z-a)^\alpha\left(\phi_0(z)+\phi_1(z)\log(z-a)+\dots+\phi_k(z)\log^k(z-a)\right)
\end{equation}
where~$\alpha$ (called \emph{an exponent} at the singularity~$a$) is a
root of the indicial polynomial and the
coefficients $\phi_i$ are power series in~$\mathbb{K}(\alpha)[[z-a]]$.
When~$a$ is an irregular singular point, then Eq.~\eqref{eq:lindeq}
admits a
basis of formal solutions of the form
\[e^{P\left(1\big/{(z-a)^{1/q}}\right)}(z-a)^\alpha\left(\phi_0(z)+\phi_1(z)\log(z-a)+\dots+\phi_k(z)\log^k(z-a)\right),\]
where~$P$ is a polynomial, $q$ a nonnegative integer and the rest as in the regular singular case, except that the power series are now in powers of~$(z-a)^{1/q}$. 
\end{theorem}
(The behavior in the neighborhood of the point~$\infty$ is obtained
from the above by changing the variable $z$ into~$1/z$ in the equation and considering~$a=0$.)

The meaning of \emph{formal} in this theorem is that these expressions
satisfy the equation formally, but no convergence to an actual
analytic solution is claimed. 
The formal aspects of this classical theory~\cite{Ince1956,CoddingtonLevinson1955,Poole1960,Wasow1987} have been transformed into computer algebra algorithms and code in the 1980's~\cite{Della-DoraTournier1981a,Tournier1987} and are now easily accessible. 
The analytic aspects are more delicate. In the regular singular case,
Frobenius showed that the power series converge in a neighborhood
of~$a$. In the
irregular
singular case, they are generally divergent. Numerical sense can still
be made of these expansions by resummation procedures~%
\cite{Balser1994,MitschiSauzin2017,Loday-Richaud2016,Delabaerre2016}.

A combination of these formal tools and those of the previous sections
forms the basis of our \emph{Dynamic Dictionary of Mathematical Functions} (DDMF)~\cite{BenoitChyzakDarrasseGerholdMezzarobbaSalvy2010}, an on-line encyclopedia\footnote{Available at \url{http://ddmf.msr-inria.inria.fr}.}
in the same spirit as the NIST DLMF\footnote{\url{https://dlmf.nist.gov}} with two major differences: only solutions of linear differential equations are handled in the DDMF and all the human expertise has been replaced by algorithms that provide an interactive access to the information, together with computer-generated proofs.

\subsection{Proofs of non-D-finiteness}
The classification of the formal behavior of solutions of linear
differential equations also provides an easy-to-use criterion to prove
that a power series is \emph{not} a solution of a linear differential
equation with polynomial coefficients, or, by passing to generating
functions, that a sequence is not the solution of a linear recurrence
with polynomial coefficients. For instance, $\tan(z)$ cannot be a
solution of such an equation, since it has infinitely many poles,
while the singularities of solutions of linear differential equations can only lie at the roots of the leading coefficient. In an analogous way, the classical Bernoulli numbers, that are present in Stirling's formula or in the Euler-Maclaurin formula, have generating function~$z/(\exp(z)-1)$ which has poles at all~$2k\pi i$, $k\in\mathbb{Z}\setminus\{0\}$ and thus cannot satisfy a linear recurrence with polynomial coefficients. Exploiting not only the number of singularities but the classification of the local behavior given above is a natural way to prove that no linear recurrence with polynomial coefficients can be satisfied by sequences \cite{FlajoletGerholdSalvy2005,FlajoletGerholdSalvy2010} like
\[\log n,\quad \sqrt{n},\quad p_n \text{ (the $n$th prime number)},\quad e^{\sqrt{n}},\quad e^{1/n}, \quad\Gamma(n\sqrt{2}),\dots\]

\subsection{Arithmetic properties}
Many generating functions~$f\in\mathbb{Q}[[x]]$ arising in
combinatorics possess the
property of being \emph{globally bounded}: $f$ has positive radius of
convergence and there exist~$a$ and~$b$ in~$\mathbb{N}\setminus\{0\}$
such that~$af(bx)\in\mathbb{Z}[[x]]$.
\begin{theorem}%
\cite{Katz1970,Andre2000a,ChudnovskyChudnovsky1985}
\label{thm:globbound}
If $F\in\mathbb{Q}[[x]]$ is differentially finite and globally
bounded, then it satisfies a Fuchsian equation (all the
singular points, including $\infty$, are regular) and all the
exponents are rational numbers.
\end{theorem}
This result also can be used to dismiss the possibility that a
given sequence
satisfies a linear recurrence. 

\begin{figure}
\centerline{\includegraphics[height=3cm]{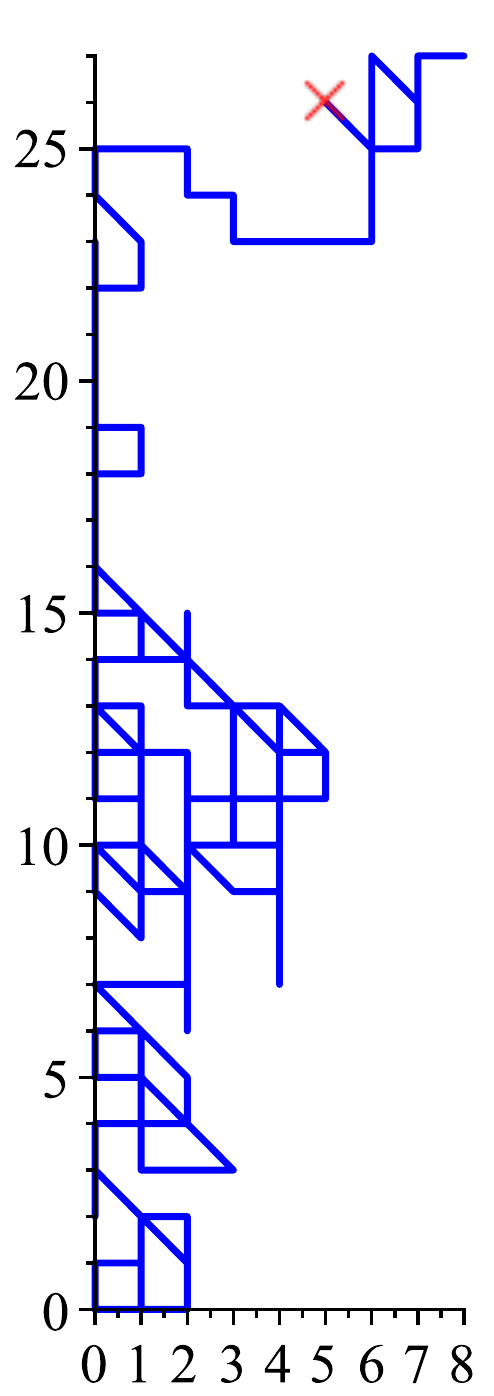}}
\caption{A walk starting from~(0,0), remaining
in~$\mathbb{N}^2$ and using 200 steps taken from $\{(-1,0),(0,1),
(1,0),
(1,-1),(0,-1)\}$. The number of such walks with $n$ steps can be
proved \emph{not} to satisfy a linear recurrence with polynomial
coefficients.\label{fig:walks}}
\end{figure}
\begin{example}
Many sequences arising in the enumeration of walks in the
quarter plane can be proved not to satisfy a linear recurrence with
polynomial
coefficients~\cite{BostanRaschelSalvy2014}. A typical example is the
number of walks on~$\mathbb{N}\times\mathbb{N}$ using~$n$ steps, all
taken in the set $\{(-1,0),(0,1),(1,0),(1,-1),(0,-1)\}$, as pictured
in Figure~\ref{fig:walks}. Using recent results connecting the
asymptotic growh of this sequence to the first eigenvalue of the
Laplacian on a spherical triangle, we obtained that this asymptotic
growth
is of the
form~$C\rho^nn^\alpha$ with $\alpha=-1+\pi/\arccos(u)$, $u$ a zero of
$8u^3-8u^2+6u-1$ so that $\alpha\not\in\mathbb{Q}$, leading to a
contradiction. 
\end{example}
\section{Singularity Analysis}\label{sec:singularityanalysis}
The asymptotic growth of a sequence~$(a_n)$ can often be analyzed
by considering its generating function
\[A(z):=\sum_{n\ge0}{a_nz^n}\]
in the complex plane. When the radius of convergence is positive, the
starting point is Cauchy's formula
\[a_n=\frac1{2\pi i}\oint{\frac{A(z)}{z^{n+1}}\,dz},\]
where the contour encloses the origin but no singularity of~$A(z)$.
\begin{figure}
\centerline{\includegraphics[height=4cm]{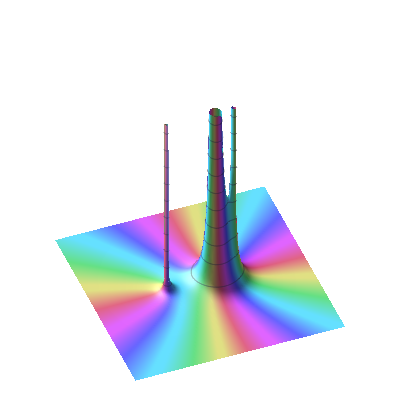}\quad
\includegraphics[height=4cm]{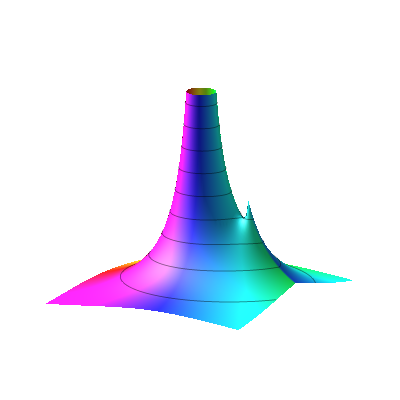}}
\caption{A view of the first Fibonacci number (left) and Catalan
number (right) in the complex plane.  (The colors indicate the
argument of the integrand.)\label{fig:cauchy}}
\end{figure}
\begin{example}
Figure~\ref{fig:cauchy} displays the absolute value of the integrand
for the cases when~$n=1$ and~$A(z)=1/(1-z-z^2)$ (left) or~$(1-
\sqrt{1-4z})/(2z)$ (right).
\end{example}
The
value at~0 is infinite due to the division by~$z^{n+1}$, which is
shown by a sort of ``chimney'' in the middle of the pictures where
the graph is truncated.
As $n$ increases, the ``chimney'' grows and the value of the integral
concentrates in a
neighborhood of the singularity of smallest modulus. This leads to a
3-step method called \emph{singularity analysis}~%
\cite{FlajoletOdlyzko1990a,FlajoletSedgewick2009}: (i) locate the
singularities of minimal modulus; (ii) compute the local behavior of
the generating function there; (iii) translate into the asymptotic
behavior of the sequence. In view of the previous two theorems, the
following is the most useful result for polynomially recursive
sequences
from combinatorics.
\begin{theorem}\cite{Jungen1931}
Let~$A(z)=\sum_{n\ge0}{a_nz^n}$ be a differentially finite power
series
with positive radius
of convergence~$\rho$. Assume that the only singularity of $A
(z)$ of modulus~$\rho$ is at $z=\rho$ and that
\[A(z)\sim c\left(1-\frac{z}\rho\right)^\alpha\log^m\frac1{1-
\frac{z}\rho},\quad z\rightarrow\rho-\]
with $\alpha\not\in\mathbb{N}$, then 
\[a_n\sim c\rho^{-n}\frac{n^{-\alpha-1}}{\Gamma(-\alpha)}\log^m
n,\quad n\rightarrow\infty.\]
\end{theorem}
Full asymptotic expansions are available as well and the case of
several singularities on the circle of convergence can be dealt with
too~\cite{FlajoletSedgewick2009}.

In the case of a polynomially recursive sequence, the linear
differential equation gives the value of~$\rho$ as a root of minimal
modulus of its leading coefficient. The computation of~$\alpha$
and~$m$ can be obtained from the differential equation as
mentioned before. The last point is the computation of the
constant factor~$c$: the initial conditions for the differential
equation are known at the origin as the first elements of the
sequence~$(a_n)$ and we need to express this solution as a linear
combination of a basis of possible behaviors at~$\rho$. In most cases,
these constants can then be obtained
numerically by analytic continuation (proving that one of the
coefficients in this linear combination is~0 is
a problem for which we only have a semi-decision algorithm). 

\begin{example}
P\'olya's random walk in~$\mathbb{Z}^d$ starts at the origin and repeatedly moves one step along one of
the axes with uniform probability. The
question is to compute the probability~$p_d$ that the walk returns to
the origin. It is a famous result of P\'olya's that~$p_2=1$. For
higher
dimension the probability is smaller than~1. Here is how it can be
computed numerically with arbitrary precision. The steps are given
for dimension~3 and that approach has been used up to dimension~15 
(where 100 digits are obtained in 1~min.):
\begin{enumerate}
	\item the probability~$u_n$ that the walk returns to the origin
	in~$2n$ steps satisfies
	\[(2n+3)(2n+1)(n+1)u_n-2(2n+3)(10n^2+30n+23)u_{n+1}+36(n+2)^3u_
	{n+2}=0\]
	(this step is not trivial);
	\item from there one could compute~$a_n:=\sum_{k=0}^n{u_k}$ which
	converges to $c:=1/(1-p_3)$, but the convergence is slow, due to a
	singularity of the generating function at~1;
	\item instead, given $a_0,a_1,a_2$, Mezzarobba's code mentioned
	above takes .4 sec. to produce 100 digits of $c,c_2,c_3$ such
	that
	\[A(z)\approx c\left(\frac{1}{1-z}+\dotsb\right)+c_2\left(
	\frac{1}{\sqrt{1-z}}+\dotsb\right)+c_3(1+\dotsb),\]
	from there, the theorem above with~$\alpha=-1,m=0$
	gives~$c$ and then~$p_3$ follows.
\end{enumerate}
In dimension~3, it turns out that a nice expression is
available~\cite{Watson1939,GlasserZucker1977}:
\[c=\frac{\sqrt{6}}{32\pi^3}
\Gamma\!\left(\frac1{24}\right)
\Gamma\!\left(\frac5{24}\right)
\Gamma\!\left(\frac7{24}\right)
\Gamma\!\left(\frac{11}{24}\right),
\]
which can be used to check our computations. In higher dimension, only
the numerical values seem available currently~\cite{HassaniKoutschanMaillardZenine2016}.
\end{example}

\section{Proofs of identities}\label{sec:proofs}
\subsection{Confinement and closure properties}
One way to prove that two power series are equal is to show that they are both solutions of a common linear differential equation, with the same initial conditions. Thus the computation is reduced to finitely many operations.
\begin{example}
Here is how one can prove that
\[\sin^2(x)+\cos^2(x)=1\]
with very little computation.

First, $\sin$ and $\cos$ are defined by a second order linear differential equation $y''+y=0$.
Next, the square of a solution to this equation is also solution of a
linear differential equation. Indeed, using the differential equation
to rewrite~$y''$ as~$-y$ shows that the~$\mathbb{Q}$-vector space
generated by~$\{y^2,yy',y'^2\}$ is closed under differentiation. Thus
if $h=y^2$, then $(h,h',h'',h''')$ are four vectors in a vector space
of dimension at most~3, which implies that they must be linearly
dependent. A linear dependency between them is precisely a linear
differential equation satisfied by~$y^2$. If needed, it is computed as
the left kernel of the matrix
\[\begin{pmatrix}
1&0&0\\
0&2&0\\
-2&0&2\\
0&-8&0
\end{pmatrix}\]
that gives the coordinates of~$(h,h',h'',h''')$ on~$(y^2,yy',y'^2)$.
This shows that $h'''+4h'=0$.
However, at this stage, it is sufficient to know that this equation
exists. Since this reasoning does not make use of the initial conditions, that same 3rd order differential equation is satisfied by~$\sin^2$ and $\cos^2$ and, by linearity, by their sum.

The constant~$-1$ is solution of a trivial first-order linear differential equation $y'=0$, so that for any~$h$ as above, $(h-1,h',h'',h''',h^{(4)})$ are five vectors in a vector space of dimension at most~4 generated by~$(-1,h,h',h'')$, implying the existence of a linear differential equation of order at most~4, with constant coefficients, satisfied by~$w:=\sin^2+\cos^2-1$. 

Now, using the initial conditions for $\sin$ and~$\cos$ to compute
\[\sin^2(x)+\cos^2(x)-1=O(x^4)\]
concludes the proof by the Picard-Lindel\"of theorem: the initial conditions defining~$w$ are~(0,0,0,0).
\end{example}
In summary, confining a power series and all its derivatives inside a
finite-dimensional vector space makes it possible to use simple linear
algebra for the proof of non-linear identities involving products
of power series. A similar reasoning applies to solutions of linear recurrences. 
\begin{example} It is a simple exercise to prove
Cassini's identity
\[F_{n+1}F_{n-1}-F_n^2=(-1)^n,\]
where $F_n$ denotes the $n$th Fibonacci number along exactly the same lines, with the recurrence~$F_{n+2}=F_{n+1}+F_n$ playing the role of $y''+y=0.$
\end{example}

With the same arguments one can prove the following classical result.
\begin{theorem}\cite[Thm.~6.4.9]{Stanley1999}\label{thm:closure-univ}
The set of power series
solutions of
linear differential equations with coefficients in~$\mathbb{K}[x]$ is a $\mathbb{K}$-algebra. So is the set of sequences solutions of linear recurrences with polynomial coefficients in~$\mathbb{K}[n]$.
\end{theorem}

\paragraph{More advanced example: Mehler's identity on the
Hermite polynomials}
\begin{equation}\label{eq:Mehler}
\sum_{n=0}^\infty{H_n(x)H_n(y)\frac{u^n}{n!}}=
\frac{\exp\left(\frac{4u(xy-u(x^2+y^2))}{1-4u^2}\right)}{\sqrt{1-4u^2}}.
\end{equation}
The starting point of the automatic proof of this identity is to
``define'' the Hermite polynomials. It will be sufficient here to use the fact that they satisfy a linear recurrence of order~2. Next, the existence of this recurrence implies that all the sequences $H_{n+k}(x)H_{n+k}(y)/(n+k)!$ for integer $k\in\mathbb{N}$ are generated over~$\mathbb{Q}(x,y,n)$ by
\[\frac{H_n(x)H_n(y)}{n!},\quad
\frac{H_{n+1}(x)H_n(y)}{n!},\quad
\frac{H_{n}(x)H_{n+1}(y)}{n!},\quad
\frac{H_{n+1}(x)H_{n+1}(y)}{n!},
\]
so that the summand in the left-hand side of Eq.~\eqref{eq:Mehler} satisfies a linear recurrence of order at most~4. That recurrence can then be translated directly into a linear differential equation satisfied by the generating function.

In that case, knowing only the order of the recurrence equation is
not sufficient anymore. Fortunately, the linear-algebra based
algorithms that compute recurrences or differential equations for sums
and products of solutions of recurrences or differential equations
have been implemented in several packages~%
\cite{SalvyZimmermann1994,Mallinger1996,KauersJaroschekJohansson2015}. \nocite{GutierrezSchichoWeimann2015}
Here, we use Maple's \texttt{gfun}.
In this computation, the $n$th Hermite polynomial in the variable~$x$ is denoted $H_x(n)$ instead of the usual $H_n(x)$. We first define the Hermite polynomials:
\begin{Verbatim}
> R[1] := {H[x](0) = 1, H[x](1) = 2*x, 
   H[x](n+2) = (-2*n-2)*H[x](n)+2*H[x](n+1)*x};
\end{Verbatim}
{\color{blue}
\[R_1 := \{H_x(0) = 1, H_x(1) = 2x, H_x(n+2) = (-2n-2)H_x(n)+2H_x(n+1)x\}\]}
\begin{Verbatim}
> R[2] := subs(x = y, R[1]);
\end{Verbatim}
{\color{blue}
\[R_2 := \{H_y(0) = 1, H_y(1) = 2y, H_y(n+2) = (-2n-2)H_y(n)+2H_y(n+1)y\}\]}
The final term of the product, $1/n!$, is defined by the recurrence $(n+1)v_{n+1}=v_n$.
Next, we compute the recurrence satisfied by the product $H_n(x)H_n(y)/n!$:
\begin{Verbatim}
> R[3] := gfun:-poltorec(H[x](n)*H[y](n)*v(n),
  [R[1], R[2], {v(n+1)*(n+1) = v(n), v(1) = 1}],
  [H[x](n), H[y](n), v(n)], c(n));
\end{Verbatim}
{\color{blue}
\begin{multline*}
R_3:=\Bigl\{  ( 16n+16 ) c (n ) -16xyc( n+1) + ( 8{x}^{2}+8{y}^{2}-8n-20)  c( n+2)\\
 -4xyc ( n+3)
 + ( n+4 ) c ( n+4) ,\\
 c ( 0 ) =1,c ( 1) =4xy, c( 2) =8{x}^{2}{y}^{2}-4{x}^{2}-4{y}^{2}+2,
 c ( 3) =\frac{32}{3}{x}^{3}{y}^{3}-16{x}^{3}y-16x{y}^{3}+24 xy,\\
\left. c ( 4 )=\frac{32}{3}{x}^{4}{y}^{4}-32{x}^{4}{y}^{2}-32{x}^{2}{y}^{4}+8{x}^{4}+96{x}^{2}{y}^{2}+8{y}^{4}-24
{x}^{2}-24{y}^{2}+6 \right\}
\end{multline*}}
The first element of that set is the recurrence, without the `$=0$' part. The other ones give the corresponding initial conditions. This recurrence is then translated into a linear differential equation for the right-hand side of Eq.~\eqref{eq:Mehler}:
\begin{Verbatim}
> gfun:-rectodiffeq(R[3], c(n), f(u));
\end{Verbatim}
{\color{blue}
\[ \left\{  ( -16{u}^{2}xy+16{u}^{3}+8u{x}^{2}+8u{y}^{2}-4
xy-4u ) f ( u ) + ( 16{u}^{4}-8{u}^{2}+1
 ) f' ( u ) ,f ( 0) =1 \right\} 
\]}
Again, the `$=0$' part is omitted from the first equation. At this stage, it is straightforward to solve this first-order equation and retrieve the desired result:
\begin{Verbatim}
> dsolve(%,f(u)) assuming 0<u,u<1/2;
\end{Verbatim}
{\color{blue}
\[ f ( u) ={\frac {{{\rm e}^{
-{\frac {4xyu-{x}^{2}-{y}^{2}}{ \left( 2u-1 \right)  \left( 2u+1
 \right) }}}}}{{{\rm e}^{-{x}^{2}-{y}^{2}}}}\sqrt {{\frac {1}{ \left( 2u+1 \right)  \left( -2u+1
 \right) }}}}
\]}

\subsection{Application to continued fractions}
Recently, we applied the same approach to the computation of explicit
formulas for continued fractions by a guess-and-prove approach~%
\cite{MaulatSalvy2015}. A typical example is provided by the continued
fraction for~$\tan z$. Starting from its definition by the Riccati
equation $y'=1+y^2$ with initial condition $y(0)=0$, it is easy to
compute the first 15 coefficients of its Taylor expansion at~0.
From there, repeatedly subtracting the first term, factoring
out the next one and inverting the rest leads to the continued fraction
\[\tan z = \cfrac{z}{1-\cfrac{z^2/3}{1-\cfrac{z^2/15}{1-\cfrac{z^2/63}{1-\cfrac{z^2/99}{1-\cfrac{z^2/143}{1-\dotsb}}}}}}.\]
From there, rational interpolation guesses automatically that the
partial numerators are given by the formula
\begin{equation}\label{eq:contfrac}
a_1(z)=z,\qquad a_n(z)=-\frac{z^2}{(2n-3)(2n-1)}\quad(n\ge2).
\end{equation}
This formula was the basis for Lambert's proof that~$\pi$ is
irrational
in~1761.

The next step is to obtain an automatic proof that the continued
fraction defined by these elements~$a_n(z)$ converges to the unique
solution of the Riccati equation with $y(0)=0$. Defining
\[H_n:=Q_n^2\left(\left(\frac{P_n}{Q_n}\right)'-1-\left(\frac{P_n}{Q_n}\right)^2\right),\]
where $P_n/Q_n$ is the $n$th convergent of the continued fraction gives a polynomial in~$P_n,Q_n,P_n',Q_n'$. A fundamental result in the theory of continued fractions is that the \emph{linear recurrence} $u_n=u_{n-1}+a_nu_{n-2}$ is satisfied by both~$P_n$ and~$Q_n$, with different initial conditions. In view of our candidate~$a_n$, we deduce that all 
$H_{n+k}$ for $k\in\mathbb{N}$ can be rewritten as linear combinations of~$P_{n+i}P_{n+j},Q_{n+i}Q_{n+j},P_{n+i}'Q_{n+j},P_{n+i}Q_{n+j}'$, for $i$ and $j$ in $\{0,1\}$. It follows that the sequence $H_n$ satisfies a linear recurrence that can be computed. The computation produces a linear recurrence of order~4 obtained without taking into account the initial conditions for~$P_n$ and~$Q_n$. Using the actual sequences makes it possible to guess the simpler
\[H_{n+1}=-\frac{z^2}{(2n+1)^2}H_n,\]
which is then proved by Euclidean division of the recurrence
operators (see~\S\ref{Ore:frac}). Thus~$H_n=O(z^{2n})$ tends to~0
as a power
series, which
concludes the proof of the formula~\eqref{eq:contfrac} without any
human intervention.

This method has been applied to all explicit C-fractions in the recent compendium by Cuyt \emph{et alii}~\cite{CuytPetersenVerdonkWaadelandJones2008}, starting from one of
\begin{itemize}
	\item a Riccati equation: $y'=A(z)+B(z)y+C(z)y^2$;
	\item a $q$-Riccati equation: $y(qz)=A(z)+B(z)y(z)+C(z)y(z)y(qz)$;
	\item a difference Riccati equation: $y(s+1)=A(s)+B(s)y(s)+C(s)y(s)y(s+1)$.
\end{itemize}
The surprising observation is that this method works in all cases, including Gauss's classical continued fraction for the quotient of contiguous hypergeometric series, its $q$-analogue due to Heine, Brouncker's continued fraction for the Gamma function. In all cases, the corresponding sequence~$H_n$ satisfies a linear recurrence of small order that is sufficient to prove the convergence. Work is in progress to explain why this method works so well and classify the formulas it yields~\cite{MaulatSalvy2018}.

\newpage
\part*{II. Conversions}
\label{sec:2}
This short second part is devoted to the middle part of Figure~%
\ref{fig:plan}:
conversions from linear differential equations to linear recurrences.
It also serves as an introduction to the operator formalism used in
the next part.

\section{Ore polynomials}
The differentiation operator $D_x$ and the operator $x$ of
multiplication by~$x$ act on power series in~$x$ and obey the
commutation law $D_xx=xD_x+1$, where~1 denotes the identity operator.
This is an operator view of the usual relation $(xf)'=xf'+f$. 

Similarly, the shift operator~$S_n$ and the operator~$n$ of
multiplication by~$n$ act on sequences indexed by~$n$, with
commutation $S_nn=(n+1)S_n$ reflecting the relation $\left.
(nu_n)\right|_{n\mapsto n+1}=(n+1)u_{n+1}$.

The analogy between these operators and polynomials has been observed
at least since the 1830s~\cite{Libri1836,Brassinne1864}. The modern
point of view was introduced by Ore a century later~\cite{Ore1931,Ore1933}.

\begin{definition} Let $\mathbb{A}$ be a ring with no zero divisor,
$\sigma$ a ring
endomorphism of~$\mathbb{A}$ and $\delta$ a $\sigma$-derivation, which
means
that for all $a,b$ in $\mathbb{A}$, $\delta(ab)=\sigma(a)\delta
(b)+\delta(a)b$.
Then the \emph{skew polynomial
ring}~$\mathbb{A}\langle\partial;\sigma,\delta\rangle$
is the ring of polynomials in~$\partial$ with coefficients
in~$\mathbb{A}$ with
usual addition and a product defined by associativity from the
commutation
\[\forall a\in\mathbb{A},\quad \partial a =\sigma(a)\partial+\delta
(a).\]
The elements of $\mathbb{A}\langle\partial;\sigma,\delta\rangle$ are called
\emph{Ore polynomials}.
\end{definition}
Special cases are the classical polynomial ring~$\mathbb{A}[x]=
\mathbb{A}\langle
x;\operatorname{Id},0\rangle$; the ring of linear differential
operators $\mathbb{K}(x)\langle D_x\rangle:=\mathbb{K}(x)\langle
D_x;\operatorname{Id},d/dx\rangle$; the ring of difference operators
$\mathbb{K}(n)\langle\Delta_n\rangle:=\mathbb{K}\langle\Delta_n;
(a(n)\mapsto a(n+1)),(a(n)\mapsto a(n+1)-a(n)\rangle$
; its close relative the ring of
recurrence
operators $
\mathbb{K}(n)\langle S_n\rangle:=\mathbb{K}(n)\langle S_n;(a(n)\mapsto
a(n+1)),0\rangle$. In
cases like this last one, where~$\delta=0$
and~$\sigma$ is
invertible, it is also natural to consider the
ring of Laurent-Ore
polynomials
in~$S_n$, denoted $\mathbb{K}(n)\langle S_n,S_n^{-1}\rangle$, with
the
obvious commutations $S_n^{-1}a(n)=a(n-1)S_n^{-1}$ and $S_nS_n^
{-1}=S_n^{-1}S_n=1$~\cite{WuLi2007}.

Ore polynomials have played an increasing role in
the design of algorithms in computer algebra since their introduction
in this area around 20 years ago~\cite{BronsteinPetkovsek1996}.

\section{Taylor morphism}
In this setting, the correspondence between linear differential
equations and linear recurrence satisfied by the sequences
of coefficients of their power series solutions becomes a ring
morphism between~$\mathbb{Q}[x,x^{-1}]\langle D_x\rangle$ and 
$\mathbb{Q}[n]\langle S_n,S_n^{-1}\rangle$, defined by
\begin{equation}\label{eq:morphism-Taylor}
D_x\mapsto (n+1)S_n,\qquad x\mapsto S_n^{-1}.
\end{equation}
(See, e.g., \cite[p.~58]{Cartier1992} for a more general statement.)

\begin{example}The Airy function~$\operatorname{Ai}(x)$ is defined
by the equation
\[y''-xy=0,\qquad y(0)=\frac{\sqrt[3]{3}}{3\Gamma(2/3)},\quad y'
(0)=-\frac{\sqrt[6]{3}\Gamma(2/3)}{2\pi}.\]
The Taylor morphism applied to differential operator~$D_x^2-x$ yields
\[D_x^2-x\mapsto(n+1)S_n(n+1)S_n-S_n^{-1}=(n+1)(n+2)S_n^2-S_n^{-1},\]
the last operator being obtained by the commutation $S_n(n+1)=
(n+2)S_n$.
This recovers the recurrence
\[(n+1)(n+2)u_{n+2}=u_{n-1}\]
from which one deduces the classical Taylor expansion
\[\operatorname{Ai}(x)=\frac{\sqrt[3]{3}}{3\Gamma(2/3)}
\sum_{n\ge0}\frac{x^{3n}}{\Gamma(n+2/3)9^nn!}
-\frac{3^{2/3}}{9}
\sum_{n\ge0}\frac{x^{3n+1}}{\Gamma(n+4/3)9^nn!}
.\]
\end{example}
\section{Chebyshev expansions}
\begin{figure}
\centerline{\includegraphics[width=.25\textwidth]{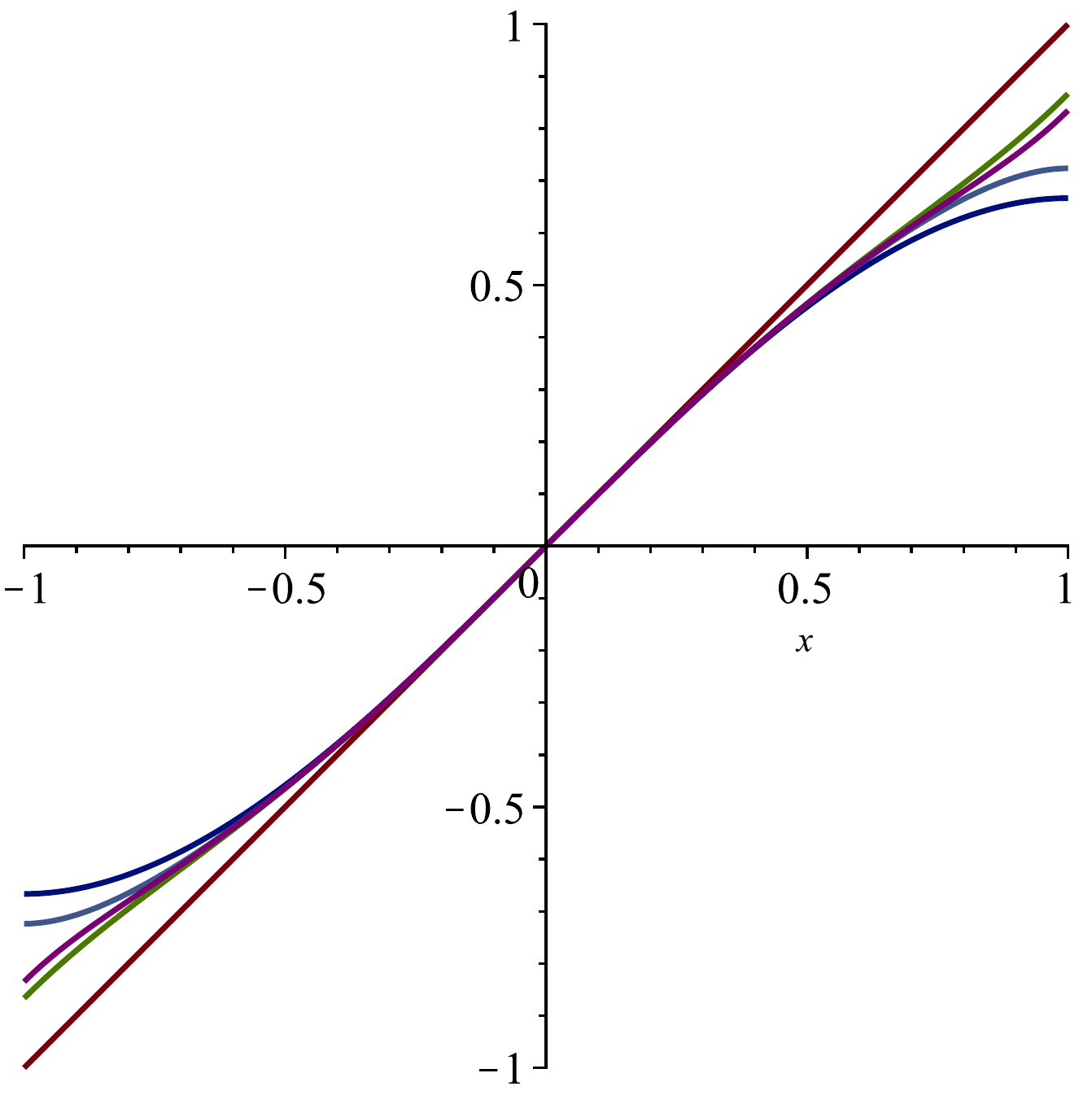}
\includegraphics[width=.25\textwidth]{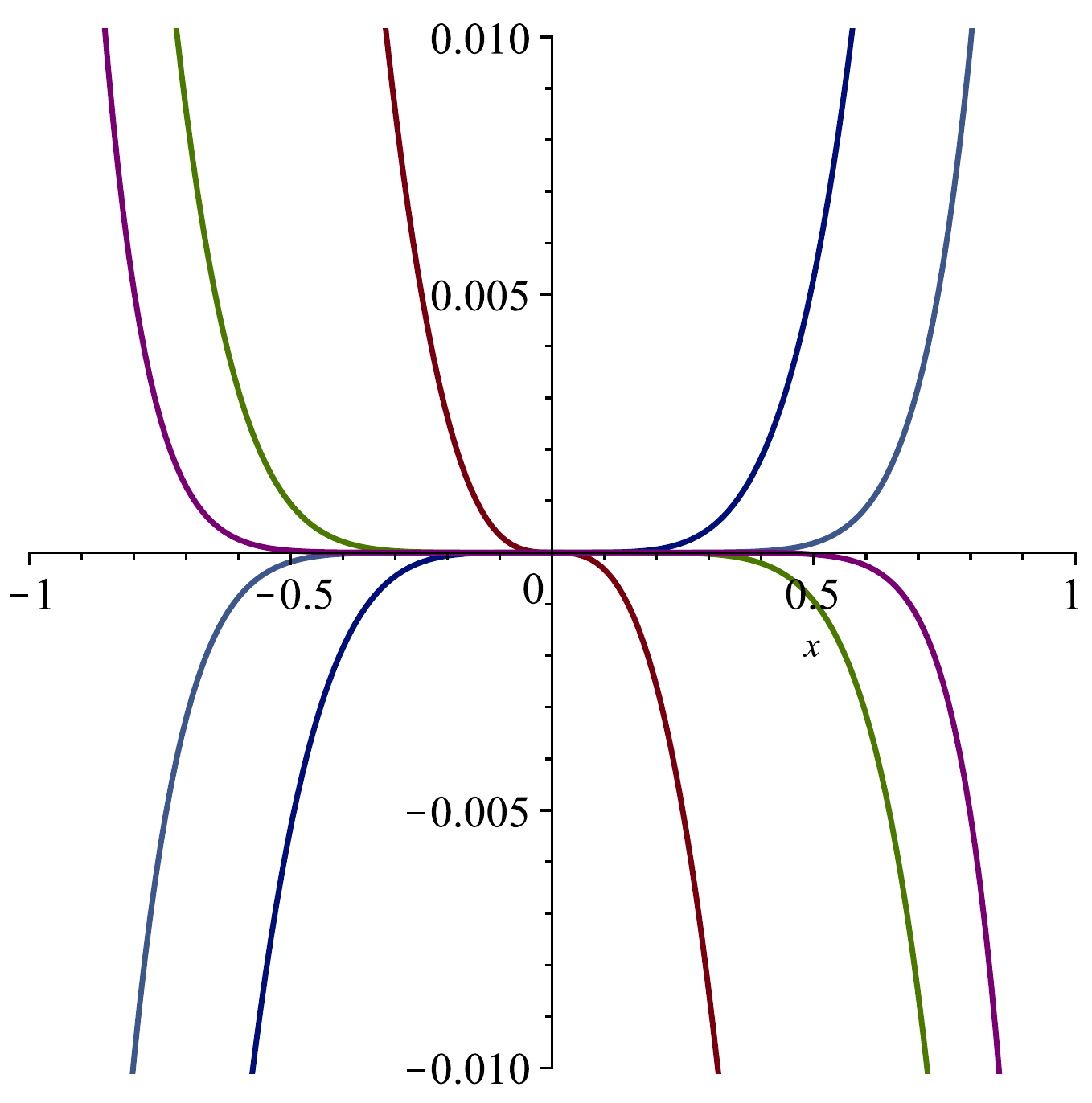}\hfill
\includegraphics[width=.25\textwidth]{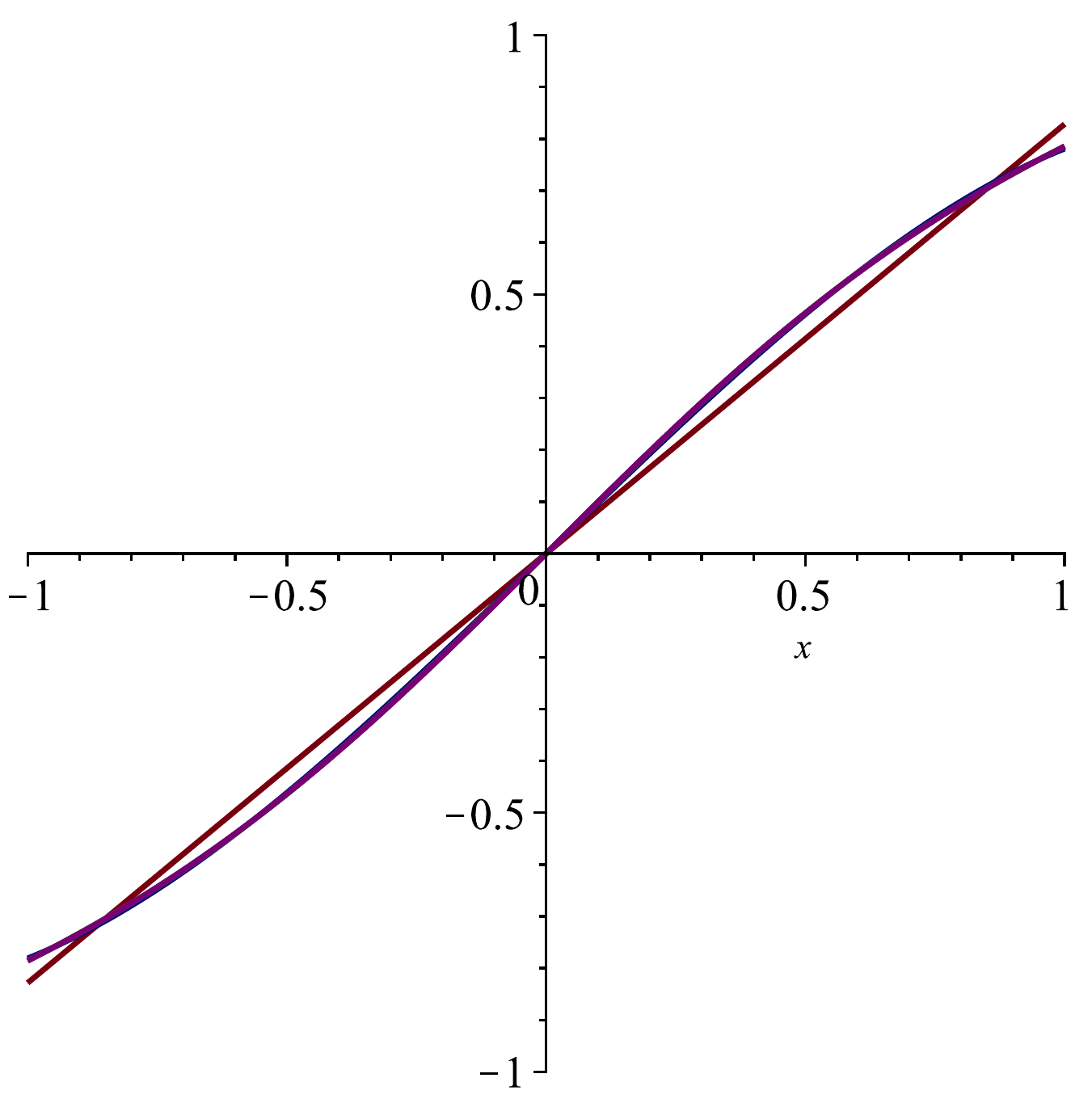}
\includegraphics[width=.25\textwidth]{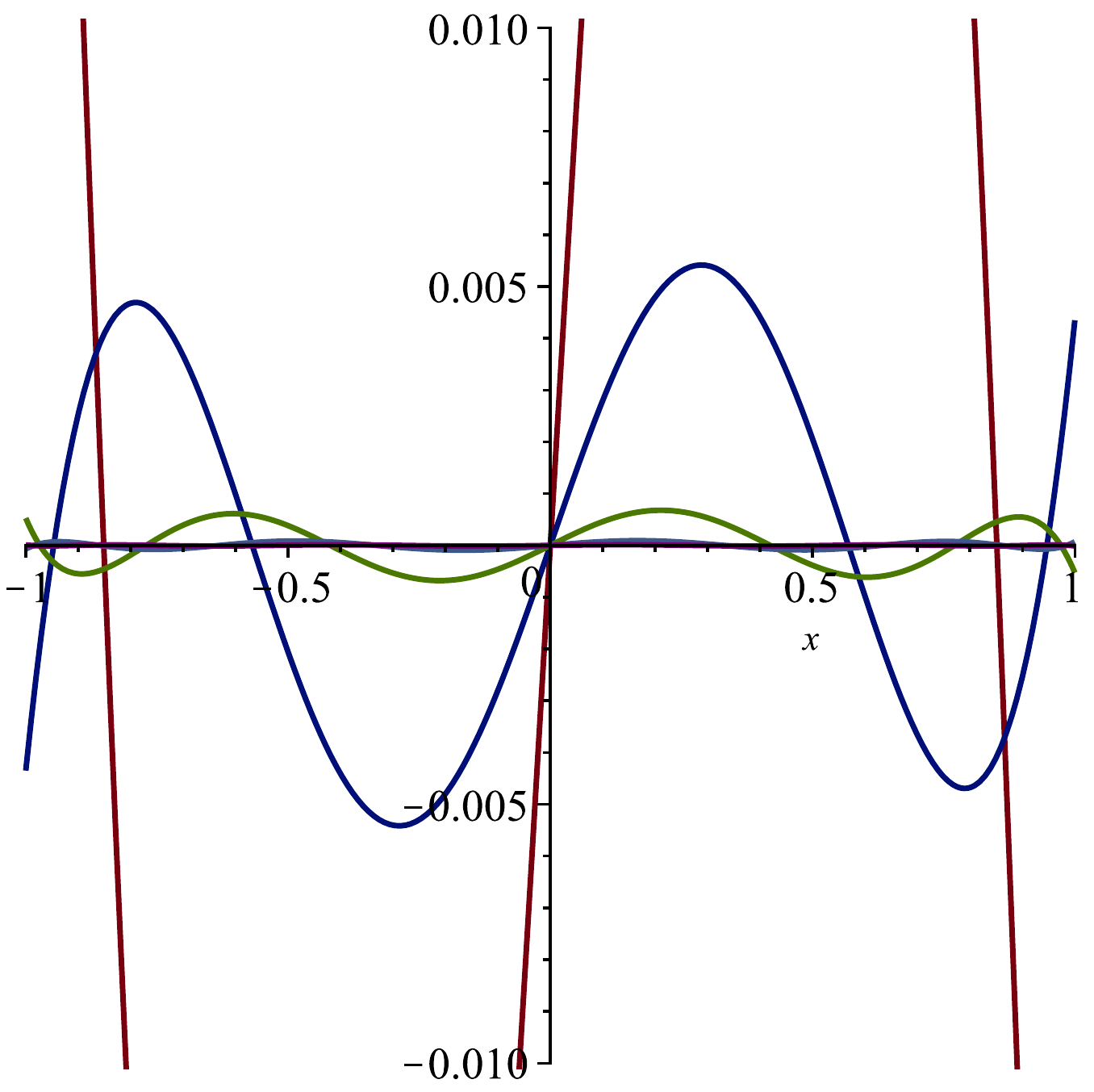}}
\caption{Truncations of Taylor expansions (left) and Chebyshev
expansions
(right) to $\arctan$, with the corresponding errors\label{fig:cheby}}
\end{figure}

Taylor expansions converge well inside their disk of convergence, but
when the aim is to approximate a function on a real interval, it is
usually preferable to use Chebyshev expansions~\cite{Cheney1998,%
MasonHandscomb2003,Trefethen2013}. This is exemplified on the case of
the function $\arctan$ on the interval~$[-1,1]$ in Figure~\ref{fig:cheby}. On
the left, the graphs of $S_n(x)$ and of $\arctan x-S_n(x)$ are
displayed
for
$n=0,\dots,4$, with
\[S_n(x)=x-\frac{x^3}{3}+\frac{x^5}{5}+\dots+(-1)^n\frac{x^{2n+1}}
{2n+1}\]
the truncation of the Taylor expansion.
On the same scale, the graphs of $C_n(x)$ and $\arctan x-C_n(x)$ are
displayed on
the right,
with
\[C_n(x)=2(\sqrt{2}-1){T_1(x)}+\dots+
\frac{(-1)^n(\sqrt{2}-1)^{2n+1}}{n+1/2}T_{2n+1}(x),
\]
where~$T_i(x)$ denotes the $i$th Chebyshev polynomial of the first
kind, defined for instance by~$T_i(\cos x)=\cos(ix)$. Already for
$C_1$, the difference with the next ones and with $\arctan$ cannot be
seen on the graph. The graphs of differences show how the error
is spread out more uniformly over the interval in the Chebyshev
expansions.

Obviously, the  
situation would be even more contrasted on an interval~$[-c,c]$
with $c>1$,
where the
Taylor expansion does not converge anymore due to the logarithmic
singularities at $\pm i$, while the Chebyshev expansion
\begin{equation}\label{eq:cheb-arctan}
\sum_{n\ge0}{\frac{(-1)^n\left(\frac{\sqrt{c^2+1}-1}{c}\right)^
{2n+1}}
{n+1/2}T_{2n+1}
(x/c)}
\end{equation}
still converges very well.

Both expansions have the property that their coefficients satisfy a
linear recurrence that can be computed automatically from the linear
differential equation. While the case of Taylor expansions uses
Ore polynomials, that of Chebyshev expansions can be computed using
Ore fractions, that we now discuss.

\section{Ore fractions}\label{Ore:frac}
By design, the degree of the product of two Ore polynomials is the sum
of their degrees. (In particular, there are no zero divisors.) Next,
for two Ore polynomials~$A$ and~$B$ \emph{with coefficients in a
field}, a right
Euclidean division~$A=QB+R$ can be defined and computed as for
commutative polynomials, except that all multiplications of~$B$ take
place on the left. From there, Euclid's algorithm for the greatest
common right divisor of $A$ and $B$ follows, as well as the extended
version
that
computes the cofactors. Moreover, performing a final iteration of this
extended Euclidean algorithm provides least common left multiples.
\begin{theorem}\cite{Ore1933}\label{thm:Ore}
Given two Ore polynomials $A$ and $B$ in a skew-polynomial
ring $\mathbb{K}\langle\partial;\sigma,\delta\rangle$ over a
field~$\mathbb{K}$, the Euclidean algorithm produces
polynomials~$u,v,G,U,V$ in~$
\mathbb{K}\langle\partial;\sigma,\delta\rangle$ such that
\[uA+vB=G,\quad UA+VB=0,\]
$G$ is a greatest common right divisor (gcrd) of~$A$ and~$B$ and $UA$
is a
least common left multiple (lclm) of them.
\end{theorem}

Now, as in the commutative case, fractions are equivalence classes of
pairs of polynomials. For our purpose, they are written with the
denominator on the left. Two fractions $B^{-1}A$ and $D^{-1}C$ are
equal when $uA=vC$ where $u$ and $v$ are such that $uB=vD=
\operatorname{lclm}(B,D)$. (Proceeding formally gives $B^{-1}A=B^
{-1}u^
{-1}uA=(uB)^{-1}uA=(vD)^{-1}vC=D^{-1}v^{-1}vC=D^{-1}C$, which
explains where this formula comes from.) It is then a simple
exercise to determine the algorithms for addition and multiplication:
\begin{align*}
B^{-1}A+D^{-1}C&=\operatorname{lclm}(B,D)^{-1}(uA+vC)\quad
&\text{where}\quad &uB=vD=\operatorname{lclm}(B,D),\\
B^{-1}AD^{-1}C&=(uB)^{-1}vC&\text{where}\quad&uA=vD=
\operatorname{lclm}(A,D).
\end{align*}
These operations turn the set of fractions into a 
(non-commutative) field~\cite{Ore1933}.

\section{Application to Chebyshev expansions}
The Taylor morphism~\eqref{eq:morphism-Taylor} is a reflection of the
action of $d/dx$ and $x$ on the basis~$(x^n)$: $(x^n)'=nx^{n-1}$ and
$x(x^n)=x^{n+1}$. Basic trigonometric identities give the analogous
relations
\begin{equation}\label{eq:defcheby}
2xT_n(x)=T_{n+1}(x)+T_{n-1}(x),\quad
2(1-x^2)T_n'(x)=-nT_{n+1}(x)+nT_{n-1}(x)
\end{equation}
for the Chebyshev polynomials. The first one indicates that $x$
 should be mapped to~$X:=
(S_n+S_n^{-1})/2$. The factor~$(1-x^2)$ in the second one prevents
such a direct translation. Proceeding formally in terms of operators
suggests that $d/dx$
should be mapped to the Ore fraction $D:=(1-X^2)^{-1}n(S_n-S_n^
{-1})/2$. Indeed, if $L(x,d/dx)$ cancels a sufficiently
smooth function~$f$, then any numerator of the Ore fraction $L(X,D)$
cancels the
coefficients of its Chebyshev expansion~\cite{BenoitSalvy2009}. This
approach sheds new light on previous algorithms in this area~%
\cite{Paszkowski1975,Lewanowicz1976,RebillardZakrajsek2006}.

\begin{example}
For~$\arctan(cx)$, A.~Benoit's package \texttt{GFS} (for Generalized
Fourier Series)~\cite{Benoit2012} produces:
\begin{Verbatim}
> deq := (c^2*x^2+1)*(diff(y(x),x,x))+2*c^2*x*(diff(y(x),x));
> diffeqToGFSRec(deq,y(x),u(n),functions=ChebyshevT(n,x));
\end{Verbatim}
{\color{blue}
\[c^2nu(n)+2(c^2+2)(n+2)u(n+2)+c^2(n+4)u(n+4)\]
}
Together with initial conditions, this leads to the formula for the
Chebyshev expansion~\eqref{eq:cheb-arctan}.
\end{example}

The numerical use of these recurrences is delicate: generally, as in
this
example, the characteristic polynomial of the leading coefficient
in~$n$, here $c^2+2(c^2+1)X+c^2X^2$, is reciprocal, which implies
that its asymptotically dominant solutions tend to infinity, while
the coefficients of Chebyshev expansions tend to~0. Thus, when
unrolling the recurrence
naively, any
numerical round-off error is eventually amplified exponentially.
Nonetheless, a recent work of
Benoit, Joldes and Mezzarobba shows how these recurrences can be
exploited, leading to an efficient
algorithm in the context of validated numerical evaluation~%
\cite{BenoitJoldesMezzarobba2017}.

\part*{III. Computing Linear Differential Equations 
(Efficiently)}
\label{sec:3}
The previous parts have shown how information can be extracted from
linear differential equations. This motivates the search of algorithms
computing linear differential equations in different contexts.

\section{Algebraic series and questions of size}
\subsection{Algebraic series can be computed fast}
A power series~$Y(X)$ with coefficients in~$\mathbb{K}$ is
called \emph{algebraic} when it is a zero of
a nonzero polynomial~$P(X,Y)\in\mathbb{K}[X,Y]$.
\begin{theorem}Algebraic power series are differentially finite.
\end{theorem}
This is an old result
that
appears in
notes of Abel's~\cite[p.~287]{Abel1992} and was
rediscovered many times~\cite{Cockle1860,Harley1862,Tannery1874}.
It implies for instance that the first $N$ coefficients
of the power series solutions of such polynomials can be computed
in~$O(N)$ arithmetic operations in~$\mathbb{K}$ (by unrolling the
recurrence).

The proof is a nontrivial but not exceedingly complicated
algorithm.
Without loss of generality, $P$ can be assumed
irreducible and we denote by~$D$ its degree. Differentiating the
polynomial equation implies
\[P_X(X,Y(X))+P_Y(X,Y(X))Y'(X)=0,\]
where~$P_X$ and~$P_Y$ denote the partial derivatives of~$P$ with
respect to~$X$ and~$Y$.
Being irreducible, $P$ is relatively prime to its
derivative~$P_Y$. Using the 
(commutative) extended Euclidean algorithm produces two
polynomials~$U$ and~$V$ in~$\mathbb{K}(X)[Y]$ such
that
\[UP_Y+VP=1.\]
This is the standard way of computing the inverse~$U$ of~$P_Y$
modulo~$P$. Denoting by~$R^{[1]}$ the remainder of the
Euclidean division
of~$-UP_X$ by~$P$ gives
\[Y'(X)=R^{[1]}(X,Y(X)),\]
with $R^{[1]}$ a \emph{polynomial} in $\mathbb{K}(X)[Y]$ of degree
in~$Y$
smaller than~$D$. Differentiating again gives
\[Y''=R^{[1]}_X+R^{[1]}_YY'
=R^{[1]}_X+R^{[1]}_YR^{[1]}=Q_2P+R^{[2]},
\]
the last term being a Euclidean division. Evaluating at~$Y(X)$ implies
that~$Y''(X)=R^{[2]}(X,Y(X))$, with $R^{[2]}$ a polynomial in $
\mathbb{K}(X)[Y]$ of degree in~$Y$ smaller than~$D$.
Iterating this process shows that all the power series~$Y^{(k)}(X)$
for $k\in\mathbb{N}$ belong to the finite-dimensional vector space
over~$\mathbb{K}(X)$ generated by~$(1,Y,\dots,Y^{d-1})$. This proves
that~$Y$ satisfies a linear differential equation of order at most~$D$
that can be obtained by linear algebra.

The same argument shows that for any~$F$ solution of a linear
differential equation and any algebraic~$Y$, $F(Y(X))$ is also
solution of a linear differential equation.

\subsection{Order-Degree curve}
\begin{figure}
\centerline{\includegraphics[width=.75\textwidth]{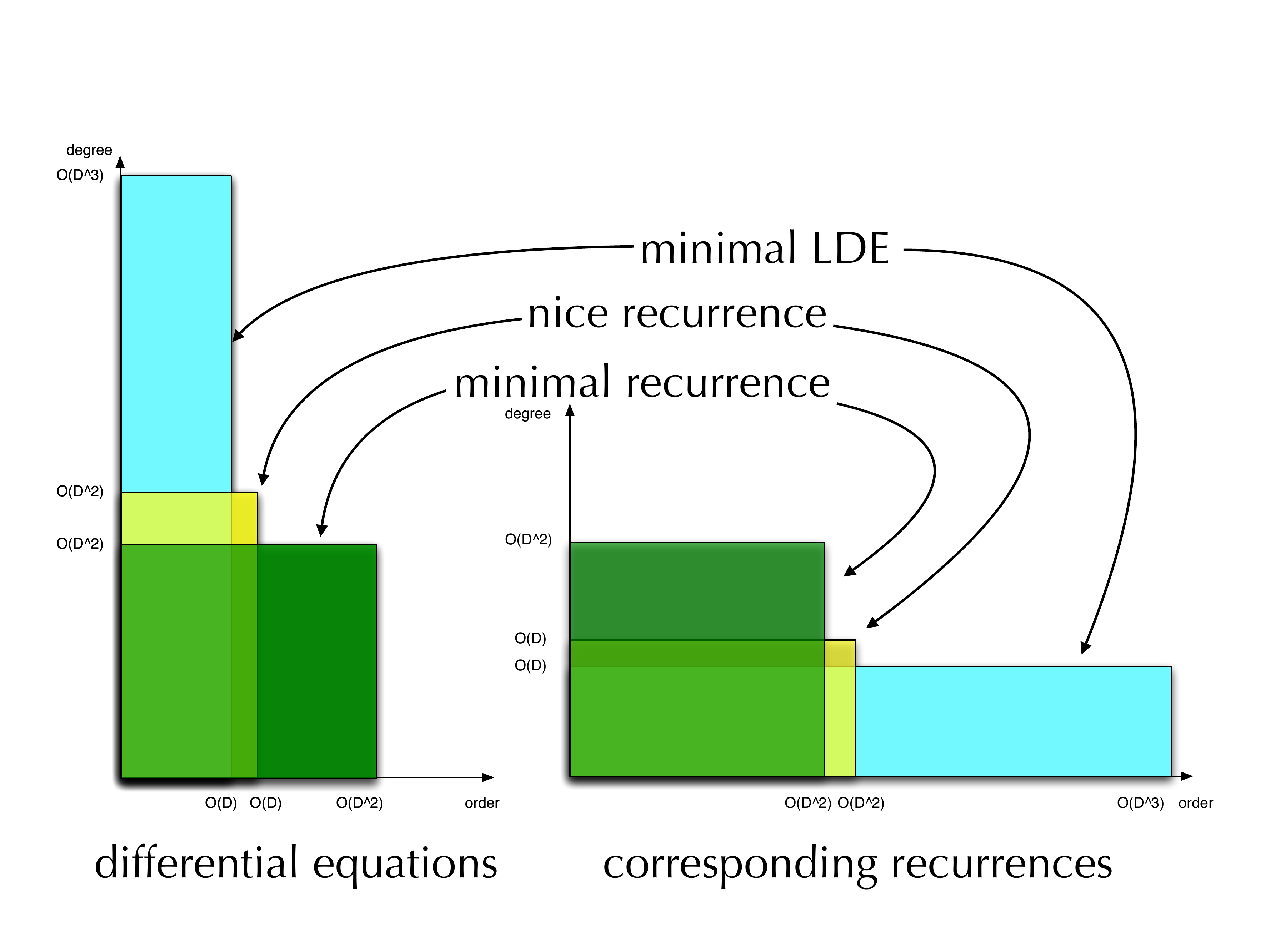}}
\caption{Differential equations and recurrences for algebraic
series\label{fig:order-degree}}
\end{figure}
The differential equation obtained by the algorithm described above
has minimal order, but the degree of its coefficients may be large. If
$D$ is also the
degree of~$P$ with respect to~$X$, then the coefficients of the
differential equation have degree~$O(D^3)$ and that bound is tight in
general.
This implies that the linear recurrence that can be deduced for the
coefficients of the power series has order~$O(D^3)$. 
Conversely, the minimal order recurrence can be shown to have order
only~$O(D^2)$ with coefficients of degree also~$O(D^2)$, thus
again, the cost of looking for minimality is a size of~$O(D^4)$
coefficients for the equation. If instead, one relaxes the constraint
on the order of the differential equation, then there always exists a
linear differential equation of order~$O(D)$ and coefficients of
degree only~$O(D^2)$~\cite{BostanChyzakLecerfSalvySchost2007}, leading to a non-minimal recurrence of
order
only~$O(D^2)$ with coefficients of degree~$O(D)$, which brings
efficiency improvements when it needs to be unrolled. These
observations are summarized in Figure~\ref{fig:order-degree}.

The large degree of the coefficients of the minimal order linear
differential equation is a general phenomenon that goes beyond the
algebraic case. It is due to the presence of numerous \emph{apparent
singularities}, that are zeros of the leading coefficient of the
differential equation, but not singularities of any of its solutions 
(e.g., $xe^x$ is a nonzero solution of a 1st order linear differential
equation, but with $y(0)=0$, which means that the Picard-Lindel\"of
theorem cannot apply at~0). Left multiples of the
differential operator let those apparent singularities disappear and
a precise analysis of the ``order-degree curve'' is possible~%
\cite{ChenJaroschekKauersSinger2013}. The apparent singularities can
all be removed algorithmically~\cite{Tsai2000a}, but the resulting
equation can have arbitrarily large order (e.g., $xy'-1000y=0$ has for
solution~$x^{1000}$ and the only way to get rid of~0 as an apparent
singularity is to go to order~1000.)
Instead, recent work has been considering ways to trade order for
degree without necessarily looking for minimal degree~%
\cite{BostanChenChyzakLi2010,ChenKauers2012}.

\section{Creative telescoping}
Creative telescoping is a method introduced by Zeilberger in the
1990s~\cite{AlmkvistZeilberger1990,Zeilberger1990,Zeilberger1991a} that computes definite integrals or sums with a free
parameter, in the sense that it produces linear differential or
recurrence equations for them. From there, the algorithms of the
previous parts can be used to compute information concerning the sum
or
the integral.
\begin{example}
Typical examples of formulas that can be computed or
proved by this method are~
\cite{Strehl1994,GrahamKnuthPatashnik1989,GlasserMontaldi1994,PrudnikovBrychkovMarichev1986a,Doetsch1930,Andrews1974b}:
\begin{gather}
\sum_{k=0}^n{\binom{n}{k}^2\binom{n+k}{k}^2}=
\sum_{k=0}^n{\binom{n}{k}\binom{n+k}{k}\sum_{j=0}^k{
\binom{k}{j}^3}},\label{eq:ex1}\\
\sum_{j,k}{(-1)^{j+k}\binom{j+k}{k+\ell}\binom{r}{j}\binom{n}
{k}\binom{s+n-j-k}{m-j}}=
(-1)^\ell\binom{n+r}{n+\ell}\binom{s-r}{m-n-\ell},\label{eq:ex2}\\
\int_0^{+\infty}{xJ_1(ax)I_1(ax)Y_0(x)K_0(x)\,dx}=\frac1{2\pi
a^2}\ln\frac1{1-a^4},\label{eq:ex3}\\
\int_{-1}^1{\frac{e^{-px}T_n(x)}{\sqrt{1-x^2}}\,dx}=
(-1)^n\pi I_n(p),\label{eq:intcheb}\\
\frac1{2\pi i}\oint{\frac{(1+2xy+4y^2)\exp\!\left(\frac{4x^2y^2}
{1+4y^2}\right)}{y^{n+1}(1+4y^2)^{3/2}}\,dy}=
\frac{H_n(x)}{\lfloor n/2\rfloor!},\label{eq:ex5}\\
\sum_{k=0}^n{\frac{q^{k^2}}{(q;q)_k(q;q)_{n-k}}}=
\sum_{k=-n}^n{\frac{(-1)^kq^{(5k^2-k)/2}}{(q;q)_{n-k}(q;q)_
{n+k}}}.\label{eq:ex6}
\end{gather}
They involve binomial coefficients, orthogonal polynomials, special
functions and their $q$-analogues. The aim of these algorithms is to
prove such identities automatically and, when the right-hand side does
not itself involve a sum or an integral, compute it from the left-hand
side. In all cases, at least one free variable remains: $n$
in~\eqref{eq:ex1}; $\ell,r,n,k,s$ in~\eqref{eq:ex2}; $a$
in~\eqref{eq:ex3}; $n$
and $p$ in~\eqref{eq:intcheb}; 
$n$ and~$x$ in~\eqref{eq:ex5}; $n$ and~$q$ in~\eqref{eq:ex6}. This
is important since the algorithms start by computing 
linear recurrences or differential equations or $q$-equations in these
free variables.
\end{example}

This part of computer algebra has made a lot of progress
in terms of generality and efficiency and is still very active. We
describe here the general context and a few of the recent
developments. More information can be found in recent surveys~%
\cite{Chyzak2014,Koutschan2013a}.

The name ``creative telescoping'' appears in van der Poorten's
enjoyable account~\cite{Van-der-Poorten1979} of Ap\'ery's proof of the
irrationality of~$\zeta(3)$. There, it was used to prove that the sum
\begin{equation}\label{eq:Apery}
A_n:=\sum_{k=0}^n{a_{n,k}},\qquad\text{with}\quad a_{n,k}=
\binom{n}{k}^2\binom{n+k}
{k}^2,
\end{equation}
satisfies the linear recurrence
\[(n+1)^3A_{n+1}-(34n^3+51n^2+25n+5)A_n+n^3A_{n-1}=0.\]
For this, an intermediate sequence 
\[b_{n,k}=4(2n+1)\left(k(2k+1)-(2n+1)^2\right)a_{n,k},\]
called \emph{the certificate} of the identity was introduced. It is
then sufficient to use simple properties of the binomial
coefficients to observe that
\[(n+1)^3a_{n+1,k}-(34n^3+51n^2+25n+5)a_{n,k}+n^3a_{n-1,k}=b_
{n,k}-b_{n,k-1}\]
and sum over~$k$, letting the right-hand side telescope.

\begin{example}
For the much simpler example of the sum 
\[U_n:=\sum_{k=0}^n{\binom{n}{k}}=(1+1)^n=2^n,\]
the computation by this method produces
\[U_{n+1}=\sum_k{\binom{n+1}{k}}=
\sum_k{\Biggl(\underbrace{\binom{n+1}{k}-\boxed{\binom{n+1}{k+1}}}+
\underbrace{\boxed{\binom{n}{k+1}}-\binom{n}{k}}+2\,\boxed{\!\binom{n}
{k}\!}\,\Biggr)}=2U_n.\]
The summands above the braces telescope and the boxed parts sum to~0
by Pascal's relation, that the method has to synthesize somehow.
\end{example}

More generally, in order to compute equations satisfied by an integral
or a sum, the method takes as input a system of equations satisfied by
the summand or integrand and relies on two operations: integration 
(resp. summation) by parts and differentiation
(resp. difference) under the integral (resp. sum) sign.
The first part gives the certificate, i.e., the multivariate expression
whose difference (or derivative) telescopes; the second part gives
the desired operator, called the \emph{telescoper}.

\section{Telescoping ideal}
Since the skew polynomial ring
$\mathbb{A}\langle\partial;\sigma,\delta\rangle$ does not have zero divisors
when $\mathbb{A}$ does not, one can iterate the construction of Ore
polynomials
and obtain multivariate Ore polynomial
rings
$\mathbb{A}\langle\partial_1;\sigma_1,\delta_1\rangle\dotsm\langle\partial_r;\sigma_r,\delta_r\rangle$.
The case when moreover $\partial_i\partial_j=\partial_j\partial_i$ for
all $
(i,j)$ is called an \emph{Ore algebra} and
denoted~$\mathbb{A}\langle\partial_1,\dots,\partial_r;\sigma_1,\dots,\sigma_r,\delta_1,\dots,\delta_r\rangle$
or even~$\mathbb{A}\langle\partial_1,\dots,\partial_r\rangle$ where the
$\sigma_i$s and $\delta_i$s are clear from the context.
If~$
\mathbb{O}$ is such an
algebra and $f$ a function on which its elements act, then the 
\emph{annihilator} of~$f$ with respect to~$\mathbb{O}$,
\[\operatorname{Ann}(f):=\{P\in\mathbb{O}\mid P(f)=0\},\]
is a left ideal in~$\mathbb{O}$. For example, the 
annihilator of $\sin x$ in~$\mathbb{Q}(x)\langle D_x\rangle$ is
generated by~$D_x^2+1$. More generally, in the case of
operators in one variable over the rational functions, the ideals are
principal by Ore's theorem
(Thm.~\ref{thm:Ore} above), thus the
annihilator of a function~$f$ is given by the greatest common right
divisor of its elements,
and rewriting on a basis of the quotient~$\mathbb{O}/\operatorname{Ann}
(f)$ is performed by Euclidean division. This is the univariate
situation considered in the previous parts.

\subsection{\texorpdfstring{$\partial$}{d}-finite ideals}
The notions of D-finiteness or P-recursiveness generalize as
follows.
\begin{definition}A left ideal $\mathcal{I}$ in a multivariate Ore
algebra~$\mathbb{O}=\mathbb{K}(\mathbf{x})\langle
\mathbf{\partial}\rangle$ is called $\partial$-finite when the
quotient~$\mathbb{O}/\mathcal{I}$ is a finite dimensional vector space
over~$\mathbb{K}(\mathbf{x})$. A function whose annihilator is
$\partial$-finite is called $\partial$-finite too.
\end{definition}
(We introduced this name with Fr\'ed\'eric Chyzak~%
\cite{ChyzakSalvy1998}, but it was probably not
such a good idea, since it is pronounced like D-finite, leading to
some confusion.) 

These ideals are a non-commutative analogue of
zero-dimensional ideals in polynomial rings. Thus, like in the
commutative case, Euclidean division and (right) gcd can be replaced
by
Gr\"obner bases~\cite{ChyzakSalvy1998}, that provide an access to a
basis of the finite-dimensional vector space~$\mathbb{O}/
\operatorname{Ann}(f)$ and to
rewriting rules reducing any element of~$\mathbb{O}/\operatorname{Ann}
(f)$ to a linear combination of the elements of this basis.
It
is
important to stress that the use of Gr\"obner bases does not raise any
efficiency issue in these computations.

\begin{figure}
\centerline{\includegraphics[height=4cm]{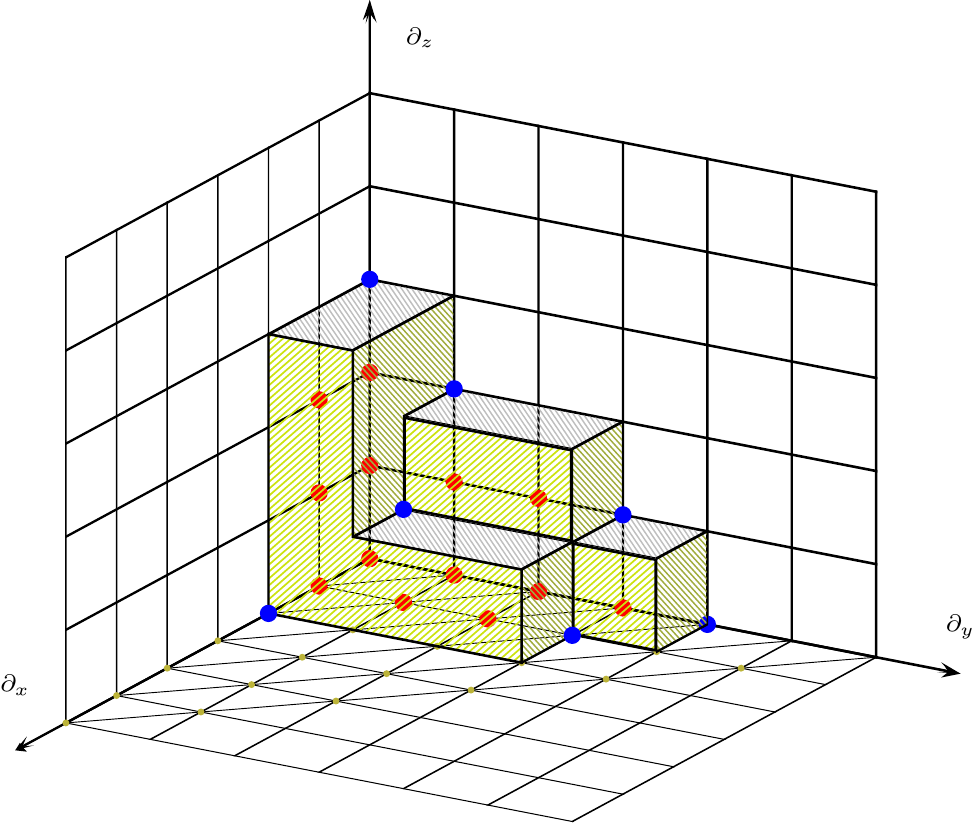}}
\caption{Illustration of Gr\"obner bases and of Chyzak's algorithm
\label{fig:chyzak}}
\end{figure}

Instead of a formal definition of the Gr\"obner basis of an
ideal~$\operatorname{Ann}(f)$, we use
Figure~\ref{fig:chyzak} to illustrate their main features. 
Each point with integer coordinates corresponds to a
monomial in~$\partial_x,\partial_y,\partial_z$ with these coordinates
as exponents. The red points, located `below' the stairs, correspond
to a basis of the quotient~$\mathbb{O}/\operatorname{Ann}(f)$. Since
the stairs are bounded, there are finitely many red points, which
shows that~$f$ is $\partial$-finite. The blue points indicate
elements of the Gr\"obner basis: each corresponds to a rewriting rule
expressing this monomial as a linear combination of the red ones. Any
monomial that is neither red nor blue is a multiple of one of the blue
ones and thus can also be reduced, possibly in several steps, to a
linear combination of the red points. 

\begin{example}
The operators defining the 
Chebyshev
polynomials of the first kind~$T_n(x)$, namely
\[(1-x^2)D_x^2-xD_x+n^2,\quad nS_{n}+(1-x^2)D_x+nx\]
make it possible to reduce any polynomial in~$\mathbb{Q}(x,n)\langle
D_x,S_n\rangle$ to a linear combination of~$1$ and~$D_x$ and
constitute a Gr\"obner basis of the ideal~$\operatorname{Ann}(T_n(x))$
is this Ore
algebra. In other words, using this basis, any $T_{n+k}^{(i)}(x)$ 
($i,k$ nonnegative
integers) rewrites as a linear combination of $T_n$ and $T_n'$, with
coefficients in $\mathbb{Q}(x,n)$.
(One could also have chosen the operators
corresponding to
the equations~\eqref{eq:defcheby}. They also give a Gr\"obner
basis in
this
algebra, for a different term order.)
\end{example}
\begin{example} Using the basis of the previous example together with
the operators
defining~$e^{-px}$, namely $(D_p+x,D_x+p)$ that form a Gr\"obner basis of~$\operatorname{Ann}(e^
{-px})$
in~$\mathbb{Q}(p,x)\langle D_p,D_x\rangle$, 
simple manipulations like those used for the proofs of univariate
identities (\S\ref{sec:proofs}) reduce to linear algebra in
finite-dimensional vector space and show that
the integrand in Eq.~\eqref{eq:intcheb} is
annihilated by the
operators
\begin{equation}\label{eq:ex-gb}
\begin{split}
D_p+x,\quad nS_n-(x^2-1)D_x-(p(1-x^2)-(n+1)x),\\
 (1-x^2)D_x^2 -(2px^2+3x-2p)D_x-(p^2x^2+3px-n^2-p^2+1),
\end{split}
\end{equation}
which constitute a Gr\"obner basis of the annihilator, showing that
the quotient in this example has dimension~2, being generated by~$1$
and~$D_x$. In other words, if~$F_n(p,x)$ denotes the
integrand of~\eqref{eq:intcheb}, all $\frac{\partial^
{i+j}}{\partial x^i\partial p^j}F_{n+k}(p,x)$ for $
(i,j,k)\in\mathbb{N}^3$ can be rewritten as linear combinations
of~$F_n$
and~$\partial F_n/\partial x$, with coefficients in $\mathbb{Q}
(n,p,x)$.
\end{example}

\subsection{Telescoping ideal}
In this framework, let the Ore algebra~$\mathbb{O}$ be $\mathbb{K}(
\mathbf{x},t)\langle\mathbf{\partial}_{\mathbf{x}},D_t\rangle$ with
$\mathbf{x}=(x_1,\dots,x_r)$ and $\mathbf{\partial}_{\mathbf{x}}=
(\partial_1,\dots,\partial_r)$ the corresponding Ore operators,
while
$D_t$ is the differentiation with respect to~$t$. If the aim is to
compute an integral of~$f$ with respect to the variable~$t$, its
representation is given by the \emph{telescoping ideal}
\[T_t(f):=\Bigl(\operatorname{Ann}(f)+
\underbrace{D_t\mathbb{K}(\mathbf{x},t)\langle
\mathbf{\partial}_{
\mathbf{x}},D_t\rangle}_{\text{int. by parts}}
\Bigr)\cap
\underbrace{\mathbb{K}(
\mathbf{x})\langle\mathbf{\partial}_{\mathbf{x}}\rangle}_{
\text{diff. under $\int$}}.\]
Indeed, canceling the derivatives that are used during the successive
integrations by parts amounts to computing modulo the \emph{right}
ideal $D_t\mathbb{K}(\mathbf{x},t)\langle
\mathbf{\partial}_{\mathbf{x}},D_t\rangle$. 
The situation in the computation of sums is completely similar, with
the differentiation operator~$D_t$ replaced by the difference
operator~$\Delta_k=S_k-1$.
\begin{example}The ideal generated by the operators in Eq.~%
\eqref{eq:ex-gb} contains
\[P=p^2D_p^2+pD_p+nD_xS_n+(px^2-nx-p)D_x+(2px-n^2-p^2-n),\]
as can be checked by reduction with the Gr\"obner basis. Rewriting
this operator as
\[P=p^2D_p^2+pD_p-(n^2+p^2)+D_x(nS_n+(px^2-nx-p))\]
shows that $p^2D_p^2+pD_p-(n^2+p^2)$ is an operator
in~$\mathbb{Q}(n,p)\langle S_n,D_p\rangle$ that belongs to the
telescoping ideal of the integrand of~\eqref{eq:intcheb} with respect
to~$x$.
\end{example}

A major source of difficulty is that while~$T_t(f)$ is a left
ideal, the sum of the left ideal $
\operatorname{Ann}(f)$ and the right ideal $D_t\mathbb{O}$ or
$\Delta_n\mathbb{O}$ is not an ideal in general, so that new
algorithms are required to perform this computation or to find
approximations (ie, subideals) of the telescoping ideal.

\paragraph{Zeilberger's slow algorithm.}
The first general approach was Zeilberger's slow algorithm~%
\cite{Zeilberger1990}, as he named it later. The idea is to restrict
integration by parts by considering only the ideal $D_t\mathbb{K}(
\mathbf{x})\langle
\mathbf{\partial}_{\mathbf{x}},D_t\rangle$. Now $D_t$ commutes with
all the elements of $\mathbb{K}(\mathbf{x})\langle
\mathbf{\partial}_{\mathbf{x}},D_t\rangle$, which makes the
computation easier. However, by restricting to a subideal,
one may be led to compute generators of much higher degree than
necessary, or even fail to find any equation. This last problem
disappears when a sufficient condition called ``holonomy'' in D-module
theory holds. Holonomy was then a starting point for Zeilberger's
approach~\cite{Zeilberger1990}.

\subsection{Towards a basis of the telescoping ideal}
Generators of the telescoping ideal can be obtained by looking for
Ore polynomials of the form
\begin{equation}\label{eq:ct}
\underbrace{\sum_{\mathbf{m}}{c_{\mathbf{m}}(\mathbf{x})
\mathbf{\partial}^{\mathbf{m}}}}_{
\text{telescoper}}+\partial_t\underbrace{\sum_{(
\mathbf{i},j)\in\mathcal{S}}{a_{\mathbf{i},j}(\mathbf{x},t)
\mathbf{\partial}^{
\mathbf{i}}\partial_t^j}}_{\text{certificate}}\in\operatorname{Ann}
(f),
\end{equation}
where, with the notations above, $\mathbf{m}=(m_1,\dots,m_r)$,
$\mathbf{i}=(i_1,\dots,i_r)$ and the multi-exponent notation is~$
\mathbf{\partial}^{\mathbf{m}}=\partial_1^{m_1}\dotsm\partial_r^
{m_r}$. In this formula, the range of the first sum is \emph{a priori}
unknown and that of the second one depends on the function~$f$
under consideration.

\paragraph{Zeilberger's fast algorithm}
Historically, the first algorithm in this family was Zeilberger's 
algorithm~\cite{Zeilberger1991a} for the definite summation of
hypergeometric sequences. These are bivariate sequences (ie, $r=1$)
whose annihilator is generated by two recurrence operators of the
form~$S_n-r(n,k)$ and~$S_k-t(n,k)$ with $r$ and $t$ rational
functions. Typical examples are the binomial coefficients or Ap\'ery's
sequence~$a_{n,k}$ from Eq.~\eqref{eq:Apery}.
Reducing any operator in~$\mathbb{O}:=\mathbb{Q}(n,k)\langle
S_n,S_k\rangle$ with
these two first-order ones leads to rational functions times the
identity. In other words, the quotient~$\mathbb{O}/\operatorname{Ann}
(f)$ is a
vector space over $\mathbb{Q}(n,k)$ of dimension~1. As a consequence,
the set of indices in the second sum of Eq.~\eqref{eq:ct} (with
$\partial_t=\Delta_k=S_k-1$) can be taken
as~$\mathcal{S}=\{(0,0)\}$ without any loss. Thus the certificate is
reduced
to one rational function. Zeilberger's algorithm takes~$m\in
\{0,\dots,r\}$ for increasing~$r$ as the set of indices for the first
sum. For each such $r$, it looks for the existence of
rational~$c_0,\dots,c_r$ and~$a_{0,0}$ by a variant of Gosper's
classical
algorithm for \emph{indefinite} summation. If a solution is found, the
algorithm stops and returns the generator of the 
telescoping
ideal~$T_k(f)$, which is
principal since this is a univariate situation.
Otherwise, the failure
to
find a solution is actually a proof that none exists and the algorithm
proceeds to the next value of~$r$. Necessary and sufficient conditions
for the algorithm to terminate are known~%
\cite{WilfZeilberger1992a,Abramov2002a,Abramov2003}. Variants of this
algorithm with quotients of dimension~1 have been developed by
Almkvist and Zeilberger~\cite{AlmkvistZeilberger1990} for integrals of
hyperexponential functions (given by two first order differential
equations) and for integrals of functions that satisfy both a first
order linear recurrence and a first order linear differential
equation.

\paragraph{Chyzak's algorithm.}
A vast generalization of Zeilberger's algorithm was designed by
 Chyzak~%
\cite{Chyzak2000} for the case when the quotient~$\mathbb{O}/
\operatorname{Ann}(f)$ is only required to have \emph{finite
dimension} over~$\mathbb{K}(\mathbf{x},t)$. 
A basis of the quotient gives the set of indices~$\mathcal{S}$ to
be used in Eq.~\eqref{eq:ct}. Then, as in Zeilberger's algorithm,
Chyzak's algorithm uses increasingly large sets of monomials with
unknown rational functions~$c_{\mathbf{m}}$ and one unknown rational
function~$a_{\mathbf{i},j}$ per element of this set~$\mathcal{S}$.
Multiplying by~$\partial_t$ on the left and reducing the resulting
expression on the basis of the quotients gives a set of linear
differential equations if $\partial_t$ is a differentiation operator 
(or recurrence equations if it is a difference operator) for these
unknown functions. The generalization of Gosper's algorithm is
replaced by algorithms for rational solutions for such systems.

Figure~\ref{fig:chyzak} suggests how the algorithm proceeds in a case
with~3 variables where integration (or summation) is performed with
respect to~$z$. During the execution of
Chyzak's algorithm, an unknown rational function~$a_{\mathbf{i},j}$ is
associated to each of the red points. The first sum in Eq.~%
\eqref{eq:ct} runs over more and more of the (small yellow) monomials
in the remaining variables~$\partial_x$ and~$\partial_y$, by
increasing order for the computation of a Gr\"obner basis of the
telescoping ideal.

\begin{example}\label{ex:16}
For the integral in Eq.~\eqref{eq:intcheb}, the
Gr\"obner basis~\eqref{eq:ex-gb} leads to considering operators of the
form
\[\sum_{(k,m)}{c_{k,m}(n,p)D_p^kS_n^m}+D_x\Big(a_0
(n,p,x)+a_1(n,p,x)D_x\Big)\]
and finding rational functions~$c_{k,m}$, $a_0$ and~$a_1$ so that they
belong to~$\operatorname{Ann}(F_n(p,x))$, or equivalently so that they
reduce to~0 by the Gr\"obner basis.

The second part of the expression does not depend on the range
of the first sum and reduces to
\[ \frac{\partial a_0}{\partial x}+a_0D_x+\frac{\partial a_1}
{\partial x}D_x+\frac{a_1}{1-x^2}\left((p^2x^2+3px-n^2-p^2_1)+
(2px^2+3x-2p)D_x\right).\]
Next, each monomial~$D_p^kS_n^m$ reduces to a linear combination~$u_
{k,m}^{(0)}+u_{k,m}^{(1)}D_x$.
Thus, by canceling the coordinates of 1 and~$D_x$ in the sum, the
problem is reduced to looking for rational solutions of the inhomogeneous linear differential system
\begin{align*}
\frac{\partial a_0}{\partial x}+\frac{a_1}{1-x^2}
(p^2x^2+3px-n^2-p^2_1)&=-\sum_{(k,m)}{c_{k,m}u_{k,m}^{(0)}},\\
\frac{\partial a_1}{\partial x}+a_0+\frac{a_1}{1-x^2}
(2px^2+3x-2p)&=-\sum_{(k,m)}{c_{k,m}u_{k,m}^{(1)}}.
\end{align*}
More precisely, the algorithm looks for values of rational~$c_{k,m}$
such that the system admits a rational solution. Several algorithms
are available for this. A solution is to: decouple the system; observe
that the poles and their multiplicities in possible rational
solutions~$a_0,a_1$ are dictated by the homogeneous part; use
undetermined coefficients on the numerator to reduce the problem to
linear algebra over the coefficients of the numerator and the~$c_
{k,m}$.

The first two cases when solutions are found is when the indices run
over the sets $\{(0,0), (0,1), (1,0)\}$ and $\{(0,0), (0,1), (0,2)\}$,
giving
\begin{equation}\label{eq:telesc-cheb}
F_{n+1}+\frac{\partial F_n}{\partial p}-\frac{n}{p}F_n=
\frac{\partial\operatorname{Cert}_1}{\partial x}
,\quad p^2\frac{\partial^2F_n}{\partial p^2}+p\frac{\partial
F_n}{\partial p}-(n^2+p^2)F_n=
\frac{\partial\operatorname{Cert}_2}{\partial x},
\end{equation}
for two explicit functions $\operatorname{Cert}_1$ and $\operatorname{Cert}_2$.
These equations can be integrated from~-1 to~1 and the left-hand
sides
provide a Gr\"obner basis of the annihilator of the integral. These
can easily be checked to cancel the Bessel function~$I_n(p)$
multiplied by~$(-1)^n$, and initial conditions can be used to conclude
the proof of Eq.~\eqref{eq:intcheb}.
\end{example}

\paragraph{Infinite dimension.}
That same method also extends to cases where the dimension of the
quotient is not finite, by proceeding by increasing total degree.
Again, termination of the algorithm is problematic, but this method
allows the automatic derivation of identities for a much larger class
of functions or sequences, including Stirling numbers, Bernoulli
numbers, the Beta function and the Hurwitz zeta function~\cite{ChyzakKauersSalvy2009}.

\paragraph{Multiple sums or integrals.}
Formally, the situation is very similar. 
The Ore algebra~$\mathbb{O}$ is $\mathbb{K}(
\mathbf{x},\mathbf{t})\langle\mathbf{\partial}_{
\mathbf{x}},\mathbf{\partial}_{\mathbf{t}}\rangle$, with
$\mathbf{x}=(x_1,\dots,x_r)$, $\mathbf{\partial}_{\mathbf{x}}=
(\partial_{x_1},\dots,\partial_{x_r})$, 
$\mathbf{t}=(t_1,\dots,t_m)$, $\mathbf{\partial}_{\mathbf{t}}=
(\partial_{t_1},\dots,\partial_{t_r})$. The aim is to
compute an integral (or sum or other depending on the Ore
operators) of~$f$ with respect to the variables~$\mathbf{t}$. The 
telescoping ideal becomes
\[T_{\mathbf{t}}(f):=\Bigl(
\operatorname{Ann}(f)+
\partial_{t_1}\mathbb{K}(\mathbf{x},\mathbf{t})\langle
\mathbf{\partial}_{\mathbf{x}},\mathbf{\partial}_{\mathbf{t}}\rangle
+\dots+
\partial_{t_m}\mathbb{K}(\mathbf{x},\mathbf{t})\langle
\mathbf{\partial}_{\mathbf{x}},\mathbf{\partial}_{\mathbf{t}}\rangle
\Bigr)\cap
{\mathbb{K}(
\mathbf{x})\langle\mathbf{\partial}_{\mathbf{x}}\rangle}.\]
Under a sufficient condition based on holonomy, Wilf and Zeilberger
have given a generalization of Zeilberger's slow algorithm and showed
that it terminates~\cite{WilfZeilberger1992}. This was improved by
Wegschaider~\cite{Wegschaider1997}.

Without restricting the integration by parts, proceeding with unknown
rational functions as above is also possible, but
it leads to a system of linear partial differential equations for
which algorithms are still missing in general. In the case of a
quotient of dimension~1, Zeilberger's fast
algorithm for hypergeometric summation has been generalized~\cite{ApagoduZeilberger2006,GaroufalidisSun2010}. Another approach for
multiple binomial sum is described below.
In the general case, except for
special
families mentioned below, one resorts to proceeding variable by
variable, with some optimizations~\cite{Chyzak2014}.

\section{Creative telescoping: new generation}\label{sec:CTnew}
The certificate computed by these algorithms is sometimes necessary:
if the integration (or summation) domain is such that the integral (or
sum) of a derivative (or a difference) is not zero, then one needs to
evaluate the certificate at the boundary of the domain. In many cases
however, it is useless. This is the case when integrating over a cycle
in~$\mathbb{C}^n$ or when summing over~$
\mathbb{Z}^n$ a product of binomial coefficients with finite
support, provided it can be ensured that the certificate does not
present singularities on the domain of integration (or summation)
that were not present in the input. However, by their design, the
algorithms
described above cannot avoid
the computation of that certificate.

\subsection{Certificates are big}
Being formed of rational functions in  more variables than the
telescoper, certificates tend to be bigger, which impacts the
complexity. 
\begin{example}The double sum
\begin{equation}\label{eq:doublesum}
C_n:=\sum_{r\ge0}\sum_{s\ge0}{(-1)^{n+r+s}\binom{n}{r}\binom{n}{s}
\binom{n+s}
{s}\binom{n+r}{r}\binom{2n-r-s}{n}}
\end{equation}
satisfies the linear recurrence
\begin{equation}\label{rec:doublesum}
(n+2)^3C_{n+2}-2(2n+3)(3n^2+9n+7)C_{n+1}-(4n+3)(4n+4)(4n+5)C_n=0,
\end{equation}
the corresponding certificate being 180kB large (approximately 2
pages of text).
\end{example}
\begin{example}
Similarly, the triple integral
\begin{equation}\label{eq:tripleint}
I(z)=\oint{\frac{(1+t_3)^2\,dt_1dt_2dt_3}{t_1t_2t_3(1+t_3
(1+t_1))(1+t_3(1+t_2))+z(1+t_1)(1+t_2)(1+t_3)^4}}
\end{equation}
satisfies the linear differential equation
\begin{multline*}
z^2(1+4z)(1-16z)I'''(z)
+3z(1-18z-128z^2)I''(z)\\
-(11-40z-444z^2)I'(z)
+2(1+30z)I(z)=0,
\end{multline*}
with a certificate that fits in 12 pages.
\end{example}
Thus, for efficiency reasons, the design of a new generation of
algorithms avoiding
the computation of the certificate has been an active research area
recently.

\subsection{Hermite reduction}
The linear system of equations obtained by reducing 
Eq.~\eqref{eq:ct} modulo the annihilator of~$f$ has a fixed
homogeneous part in the unknown rational coefficients~$a_{
\mathbf{i},j}$ and a variable inhomogeneous part coming from the
telescoper. The idea of algorithms based on Hermite reduction is to
work modulo the image of the linear map constituted by the homogeneous
part. When a finite basis of the quotient by this image is available,
generalized Hermite reduction is the process of reducing (vectors
of) rational functions to this basis. This generalizes the classical
Hermite reduction, which reduces modulo the image of a
derivation~$D_x$.

This was first exploited in the case of dimension~1 for bivariate
rational functions~\cite{BostanChenChyzakLi2010}, for hyperexponential
functions~\cite{BostanChenChyzakLiXin2013a}, for bivariate
hypergeometric terms~\cite{ChenHuangKauersLi2015,Huang2016}, for mixed
hypergeometric-hyperexponential functions~%
\cite{BostanDumontSalvy2016}. Next, it was extended to algebraic
functions~\cite{ChenKauersSinger2012,ChenKauersKoutschan2016}, to
Fuchsian functions~\cite{ChenHoeijKauersKoutschan2018}, to solutions
of differential systems~\cite{Van-Der-Hoeven2017} and finally to
the integration of $\partial$-finite
functions~\cite{BostanChyzakLairezSalvy2018}.

A further simplification is brought by the use of adjoint operators.
If~$L=c_rD_x^r+\dots+c_0\in\mathbb{K}
(x)\langle D_x\rangle$, then its adjoint is defined as~$L^*=c_0+\dots+
(-D_x)^rc_r$. It is related to integration by parts via Lagrange's
identity
\[uL(f)-L^*(u)f=D_x(P_L(f,u)),\]
satisfied for any $u$ and $f$, with an explicit~$P_L$. Thus, if $f$ is
a solution of~$L$, any
rational function~$R$ in~$L^*(\mathbb{K}(x))$ is such that~$Rf$ is a
derivative. Now, if, as in the case of Example~\ref{ex:16}, all the
other operators in the algebra rewrite as linear combinations of
powers of~$D_x$ (see Eq.~\eqref{eq:ex-gb}), then all operations
boil down to Hermite reductions of rational functions. This specific
form can always be achieved by the use of a so-called \emph{cyclic
vector}~\cite{ChurchillKovacic2002}.
\begin{example}
The adjoint of the last operator in the basis~\eqref{eq:ex-gb} is
\[M:=(x^2-1)D_x^2+(x-2p(x^2-1))D_x+(p^2(x^2-1)-px-n^2).\]
If one wants to reduce a polynomial with respect to~$M$, the first
step is to determine the intersection of~$M(\mathbb{Q}(x))$ with
$\mathbb{Q}[x]$. Considering~$M(x^k)$ for $k\in\mathbb{N}$ shows
that all polynomials of degree at least~2 belong to~$M(
\mathbb{Q}(x))\cap\mathbb{Q}[x]$. To prove that no other polynomial
belong to this set, it is sufficient to consider the singularities
at~$\pm1$ and observe that $M$ increases the orders of the poles
there. Thus, 1 and~$x$ reduce to
themselves with
respect to~$M$ and the Hermite reduction of any polynomial is a
linear combination of~$1$ and~$x$ with
coefficients in~$\mathbb{Q}(n,p)$. In particular, using $M(1)$ reduces $x^2$ to $x/p+1+n^2/p^2$.

This means first that~$F_n$ itself is not a derivative (or 1 would be
reduced to~0), that no linear combination of~$F_n$ and~$\partial
F_n/\partial p$ is a derivative (since~1 and~$x$ are linearly
independent). Next, $D_p$ reduces to~$x$ by the Gr\"obner basis,
so~$D_p^2$ reduces to~$x^2$ and the
Hermite reduction of~$x^2$ implies that
\[p^2\frac{\partial^2F_n}{\partial p^2}+p\frac{\partial
F_n}{\partial p}-(n^2+p^2)F_n\]
is a derivative, which recovers the second part of Eq.~%
\eqref{eq:telesc-cheb}. Finally, rewriting the equation for~$F_{n+1}$
in the Gr\"obner basis~\eqref{eq:ex-gb} by a Euclidean right division
by~$D_x$ gives
\[nS_n-D_x(x^2-1)+(px^2+(n-1)x-p),\]
so that again, the Hermite reduction of~$x^2$ helps conclude that
\[F_{n+1}+\frac{\partial F_n}{\partial p}-\frac{n}{p}F_n\]
is a derivative, which is the first part of Eq.~%
\eqref{eq:telesc-cheb}, obtained
without computing the certificates.
\end{example}

\subsection{Periods}\label{sec:period}
Integrals of rational functions over cycles provide 
an important class of \emph{multiple} integrals where the computation
of the certificate
is
unnecessary. 
What we call \emph{period} here is an
integral of a rational
function in~$\mathbb{Q}(\mathbf{t})$ with $\mathbf{t}=
(t_1,\dots,t_m)$ over a cycle in~$
\mathbb{C}^m$ that avoids the zero-set of the denominator. These
numbers form an important subclass of the countable class of periods
considered
by
Kontsevich and Zagier~\cite{KontsevichZagier2001}, with fewer
constraints on the domain of integration. 

If instead one integrates in~$\mathbb{C}^m$ a function~$F$ in~$
\mathbb{Q}
(x,\mathbf{t})$ for
an extra variable~$x$ and if the denominator does not vanish in a
neighborhood of the cycle of integration, then the period is a
function of~$x$. 
Moreover, this function satisfies
a linear differential equation, called a \emph{Picard-Fuchs
equation} after early work
by Picard~\cite{Picard1902} in the bivariate case.

Without loss of generality, $F\in\mathbb{Q}
(x,\mathbf{t})$ can be written~$P/Q^\ell$ with $Q$ a square-free
polynomial. An algorithm finding the Picard-Fuchs equation is
obtained by a process
called Griffiths-Dwork reduction, which can be seen as a
generalization of Hermite's reduction~%
\cite{Griffiths1969,Dwork1964,Christol1985}. A first technicality is that
in order to get a better
control over the degrees, one homogenizes the integrand by
introducing a new
variable~$t_0$. Next, a key step is to introduce the 
ideal
generated by the partial derivatives~$\partial_0Q,\dots,\partial_mQ$.
The reduction takes the remainder modulo (a Gr\"obner basis of) this
ideal of
the numerators that appear and use
integration by parts: if $P=r+v_0\partial_0Q+\dots+v_m\partial_mQ$
and~$\ell>1$,
then
\[\frac{P}{Q^\ell}=\frac{r}{Q^\ell}-\frac{1}{\ell-1}\left
(\partial_0\frac{v_0}{Q^{\ell-1}}+\dots+\partial_m\frac{v_m}
{Q^{\ell-1}}\right)
+\frac{1}{\ell-1}\frac{\partial_0v_0+\dots+\partial_mv_m}{Q^{\ell-1}}.
\]
Thus, modulo derivatives, $P/Q^\ell$ reduces to~$r/Q^\ell$ and a
rational function with denominator only~$Q^{\ell-1}$ on which the
process is
repeated until~$\ell=1$ is reached. A result of Griffiths~%
\cite{Griffiths1969} shows
that,
under some regularity condition, $F$ is reduced  to~0 by this process
if and only if the integral of~$F$ over cycles is~0. The computation
of the Picard-Fuchs equation then consists in computing the reductions
of the successive derivatives with respect to the free variable~$x$
and looking for a
linear relation between the reductions, whose coefficients are those of
the differential equation. When the regularity conditions are not
met, they can be recovered by a perturbation method~\cite{Dwork1964}.
Counting dimensions
carefully and using recent efficient algorithms for the reduction
stage leads to the following~\cite{BostanLairezSalvy2013,Lairez2016}.
\begin{theorem}\label{thm:BostanLairezSalvy2013}
Let~$F=P/Q$ be a rational function in $\mathbb{Q}
(x,\mathbf{t})$ with $\mathbf{t}=(t_1,\dots,t_m)$, let
\[N=\max(\deg_\mathbf{t}P+m+1,\deg_\mathbf{t}Q)\quad\text{and}\quad
d_x=\max
(\deg_xP,\deg_xQ).\]
Then $F$ admits a telescoper whose certificate is singular only
where~$Q=0$. This telescoper has order at most~$N^m$ and degree $O
(N^{3m}d_x)$. It can be computed in~$O(N^{8m}d_x)$ arithmetic
operations in~$\mathbb{Q}$.
\end{theorem}
The bound on the order is tight. It is important to note that
generically, the certificate has a number of monomials growing
like~$N^{n^2/2}$ and thus cannot even be written within that
complexity.

Recent work has exploited these differential equations for 
the computation of volumes of semi-algebraic sets~%
\cite{LairezSafey-El-Din2018} and of multiple binomial sums (see
below).

\section{Diagonals} Diagonals form an important class of such multiple
integrals of rational functions.
If $F(\mathbf{t})=G(\mathbf{t})/H(\mathbf{t})$ with~$\mathbf{t}=
(t_1,\dots,t_m)$ is a multivariate rational function such that~$H
(0)\neq0$, then it admits a Taylor expansion
\[F(\mathbf{t})=\sum_{\mathbf{i}\in\mathbb{N}^m}{c_
\mathbf{i}\mathbf{t}^{\mathbf{i}}}\]
and its
\emph{diagonal} is the power series
\[\Delta F(t):=\sum_{k\in\mathbb{N}}{c_{k,k,\dots,k}t^k}.\]
\begin{example}
The simplest example is the diagonal of Pascal's triangle: the
binomial coefficients
are the Taylor coefficients of~$f=1/(1-x-y)$ and the central binomial
coefficients~$\binom{2k}{k}$ have for generating function $\Delta
f$. Less obvious are
\begin{align*}
\sum_{k=0}^\infty{\frac1{k+1}\binom{2k}{k}t^k}&=\Delta\frac{1-2x}{
(1-x-y)(1-x)},\\
\sum_{k=0}^\infty{A_kt^k}&=\Delta\frac1{1-t(1+x)(1+y)(1+z)
(1+y+z+yz+xyz)},
\end{align*}
where the first one is the generating function of the Catalan numbers
and the second one is that of the Ap\'ery numbers from Eq.~%
\eqref{eq:Apery}.
\end{example}
Since diagonals can be rewritten as multidimensional
residues 
\[\Delta F(t)=\left(\frac1{2\pi i}\right)^{m-1}\oint{F\!\left
(t_1,\dots,t_{m-1},\frac{t}{t_1\dotsm t_{m-1}}\right)\frac{dt_1\dotsm
dt_{m-1}}{t_1\dotsm t_{m-1}}},\]
the results of the previous section apply and lead to the following.
\begin{theorem}\cite{Christol1984}
	Diagonals of rational functions are differentially finite.
\end{theorem}
Moreover, if~$F$ has degree~$d$, then, by Theorem~%
\ref{thm:BostanLairezSalvy2013}, the
differential
equation satisfied by the diagonal has order that grows like~$d^m$ and
its
coefficients have degree bounded by~$d^{O(m)}$. It can be computed in
good complexity.

Much more is known about diagonals. Algebraic series  
are the
diagonals
of bivariate rational functions~%
\cite{Polya1921,Furstenberg1967} (the
degree of the polynomial may be large~\cite{BostanDumontSalvy2017});
diagonals are closed under sum, product and Hadamard product.
They are globally bounded
and therefore satisfy the hypothesis of Theorem~\ref{thm:globbound}; 
Christol conjectures that the converse holds: all globally bounded
D-finite power series would be diagonals.
More information
on diagonals can be found in recent surveys~%
\cite{BostanBoukraaChristolHassaniMaillard2013,Christol2015}.

Also, for the most regular of those rational functions, the constant
involved in the asymptotic behavior of the coefficients of their
diagonals, as discussed
in~\S
\ref{sec:singularityanalysis}, can sometimes be computed
explicitly and moreover algorithmically~%
\cite{PemantleWilson2013,MelczerSalvy2016}.

\section{Multiple binomial sums} These are sums like that of Eq.~%
\eqref{eq:doublesum}. A more formal definition is the following.
\begin{definition} The class of \emph{multiple binomial sums}  over~$\mathbb{K}$ is the
class of sequences of elements of~$\mathbb{K}$ obtained from: geometric
sequences $n\mapsto C^n$ (for $C\in\mathbb{K}\setminus\{0\}$),
binomial coefficients $(n,k)\mapsto\binom{n}{k}$, the Kronecker delta
sequence~$n\mapsto\delta_n$ (which is 1 at index~$n$ and 0 everywhere
else) using the operations of: addition, multiplication,
multiplication by a scalar, affine change of indices $u_{
\mathbf{n}}\mapsto u_{\Lambda\mathbf{n}}$ with~$\Lambda$ an affine map
from~$\mathbb{Z}^d$ to~$\mathbb{Z}^e$ and indefinite summation 
\[(\mathbf{m},n)\mapsto\sum_{k=0}^n{u_{\mathbf{m},k}}.\]
\end{definition}
These sums are very closely related to diagonals, by the following
not too difficult result, whose proof is  effective.
\begin{theorem}\cite{BostanLairezSalvy2017} A sequence $u:
\mathbb{N}\rightarrow\mathbb{K}$ is a multiple binomial sum if and
only if the generating function~$\sum_{n\ge0}u_nt^n$ is the diagonal
of a rational power series.
\end{theorem}
In order to compute a linear recurrence for a multiple binomial sum,
it is actually not necessary to rewrite it as a diagonal,
and a residue expression is sufficient. This provides a fast
algorithm for
single or multiple summation~\cite{BostanLairezSalvy2017} that makes
effective a classical approach sometimes called the generating
function method~\cite{Egorychev1984}.
\begin{example}
Dixon's classical identity
\[\sum_{k=0}^{2n}{(-1)^k\binom{2n}{k}^3}=(-1)^n\frac{(3n)!}{n!^3}\]
is computed automatically by first expressing the generating function
of the sum as the integral of a rational function as follows. A
starting point is to define~$\binom{n}{k}$ as the
coefficient
of~$x^k$ in~$(1+x)^n$, hence, by Cauchy's formula, as
\[\binom{n}{k}=\frac1{2\pi i}\oint{\frac{(1+x)^n}{x^k}
\frac{dx}{x}},\]
where the contour is a small circle (of radius smaller than~1) around
the origin. Then the summand has for integral representation
\[(-1)^k\binom{2n}{k}^3=
\frac1{(2\pi i)^3}\oint{\left(\prod_{i=1}^3{
(1+x_i)^2}\right)^n\left(\frac{-1}{x_1x_2x_3}\right)^k}
\frac{dx_1dx_2dx_3}{x_1x_2x_3},\]
where the contour is the product of three of those small circles.
Multiplying by~$t^n$ and summing the geometric series over~$k$ and~$n$
finally gives the generating function of the sum as
\[\frac1{(2\pi i)^3}\oint{\frac{x_1x_2x_3-t\prod_{i=1}^3{
(1+x_i)^2}}{\left(x_1^2x_2^2x_3^2-t\prod_{i=1}^3{
(1+x_i)^2}\right)\left(1-t\prod_{i=1}^3{
(1+x_i)^2}\right)}\,dx_1dx_2dx_3}.\]
Next, the algorithm detects that the integral with respect to one of
the variables, say~$x_3$, can be obtained by residue computation,
taking into account that as~$t\rightarrow0$, the first factor of the
denominator has all its roots that remain small, while those of the
second one do not contribute. The integral is thus simplified to
\[\frac1{(2\pi i)^2}\oint{\frac{x_1x_2\,dx_1dx_2}{x_1^2x_2^2-t
(1+x_1)^2(1+x_2)^2(1-x_1x_2)^2}}.\]
From there, the algorithms of \S\ref{sec:period} produce the following
linear differential equation for the generating function:
\[t(1+27t)y''+(1+54t)y'+6y=0,\]
which in turn gives the linear recurrence
\[3(3n+2)(3n+1)u_n+(n+1)^2u_{n+1}=0,\]
concluding the proof of Dixon's formula after checking one
initial condition. Actually, the right-hand side is discovered
automatically by this computation.
\end{example}
\begin{example}
From the double sum
from Eq.~\eqref{eq:doublesum}, the rational function integrand of
Eq.~\eqref{eq:tripleint} is obtained automatically. From there, the
Picard-Fuchs equation is deduced and then by direct translation into a
recurrence, Eq.~\eqref{rec:doublesum} follows.
\end{example}
And again, from the linear differential equation or linear recurrence,
a lot of information can be obtained for the sum by the methods of the
first part. 


\bibliographystyle{spmpsci-fixdoi}      

\end{document}